\documentclass{article}
\usepackage{soul}
\usepackage[table]{xcolor}  
\usepackage{cite}
\usepackage{microtype}
\usepackage{graphicx}	
\usepackage{booktabs} 
\usepackage{colortbl}
\usepackage{longtable}
\usepackage{subcaption}
\usepackage{hyperref}

\usepackage[arxived]{icml2024}

\usepackage{amsmath}
\usepackage{amssymb}
\usepackage{mathtools}
\usepackage{amsthm}
\usepackage{threeparttable}
\usepackage{xspace}

\usepackage{paralist}
\usepackage{multirow}

\usepackage{listings} 

\definecolor{codegreen}{rgb}{0,0.6,0}
\definecolor{codegray}{rgb}{0.5,0.5,0.5}
\definecolor{codepurple}{rgb}{0.58,0,0.82}
\definecolor{backcolour}{rgb}{0.95,0.95,0.92}

\usepackage[capitalize,noabbrev]{cleveref}

\theoremstyle{plain}

\theoremstyle{definition}

\theoremstyle{remark}

\usepackage[textsize=tiny]{todonotes}

\usepackage{pifont}

\newif\ifshowcomments
\showcommentsfalse

\newcommand{\mytodogreen}[1]{\textcolor{green}{\ding{46}~{\sf}~#1}}
\newcommand{\mytodoorange}[1]{\textcolor{orange}{\ding{46}~{\sf}~#1}}
\newcommand{\mytodocyan}[1]{\textcolor{cyan}{\ding{46}~{\sf}~#1}}
\newcommand{\mytodopink}[1]{\textcolor{purple}{\ding{46}~{\sf}~#1}}

\ifshowcomments
\newcommand{\gang}[1]{\mytodogreen{[gang: #1]}}
\newcommand{\cp}[1]{\mytodogreen{[chengpeng: #1]}}
\newcommand{\xjlin}[1]{\mytodoorange{[xjlin: #1]}}
\newcommand{\xiaoheng}[1]{\mytodocyan{[xiaoheng: #1]}}
\newcommand{\pengd}[1]{\mytodopink{[pengd: #1]}}
\newcommand{\lming}[1]{\mytodopink{[lming: #1]}}
\newcommand{\youheng}[1]{\mytodopink{[youheng: #1]}}
\newcommand{\zgh}[1]{\mytodopink{[zgh: #1]}}

\else
\newcommand{\gang}[1]{}
\newcommand{\cp}[1]{}
\newcommand{\xjlin}[1]{}
\newcommand{\xiaoheng}[1]{}
\newcommand{\pengd}[1]{}
\newcommand{\lming}[1]{}
\newcommand{\youheng}[1]{}
\newcommand{\zgh}[1]{}

\fi

\newcommand{\ToolName}{\textsc{RepoFuse}\xspace}

\newcommand{\GraphName}{\textsc{Code Knowledge Graph}\xspace}

\icmltitlerunning{\ToolName: Repository-Level Code Completion with Fused Dual Context}

\begin{document}

\twocolumn[
\icmltitle{\ToolName: Repository-Level Code Completion with Fused Dual Context}




\icmlsetsymbol{equal}{*}
\begin{icmlauthorlist}
\icmlauthor{Ming Liang}{antgroup}
\icmlauthor{Xiaoheng Xie}{antgroup}
\icmlauthor{Gehao Zhang}{antgroup}
\icmlauthor{Xunjin Zheng}{antgroup}
\icmlauthor{Peng Di}{antgroup}
\icmlauthor{Wei Jiang}{antgroup}
\icmlauthor{Hongwei Chen}{antgroup}
\icmlauthor{Chengpeng Wang}{antgroup}
\icmlauthor{Gang Fan}{antgroup}
\end{icmlauthorlist}

\icmlaffiliation{antgroup}{Ant Group, China}

\icmlcorrespondingauthor{Gang Fan}{fangang@antgroup.com}
\icmlcorrespondingauthor{Wei Jiang}{jonny.jw@antgroup.com}

{\theoremstyle{definition}\newtheorem{researchquestion}{RQ}}
{\theoremstyle{definition}\newtheorem{casestudy}{Case Study}}
{\theoremstyle{definition}\newtheorem{problem}{Problem}}
{\theoremstyle{definition}\newtheorem{definition}{Definition}}
{\theoremstyle{definition}\newtheorem{example}{Example}}

{\theoremstyle{definition}\newtheorem{prompt}{Prompt}}
{\theoremstyle{definition}\newtheorem{response}{Response}}

\vskip 0.3in
]



\printAffiliationsAndNotice{}  

\begin{abstract}
The success of language models in code assistance has spurred the proposal of repository-level code completion as a means to enhance prediction accuracy, utilizing the context from the entire codebase. However, this amplified context can inadvertently increase inference latency, potentially undermining the developer experience and deterring tool adoption—a challenge we termed the \emph{Context-Latency Conundrum}.
This paper introduces \ToolName, a pioneering solution designed to enhance repository-level code completion without the latency trade-off. \ToolName uniquely fuses two types of context: the \emph{analogy context}, rooted in code analogies, and the \emph{rationale context}, which encompasses in-depth semantic relationships. We propose a novel \emph{rank truncated generation (RTG)} technique that efficiently condenses these contexts into prompts with restricted size. This enables \ToolName to deliver precise code completions while maintaining inference efficiency. Through testing with the CrossCodeEval suite, \ToolName has demonstrated a significant leap over existing models, achieving a 40.90\% to 59.75\% increase in exact match (EM) accuracy for code completions and a 26.8\% enhancement in inference speed. Beyond experimental validation, \ToolName has been integrated into the workflow of a large enterprise, where it actively supports various coding tasks.


\end{abstract}

\section{Introduction}
\label{introduction}

Language models (LMs) demonstrate exceptional skill in a variety of programming tasks, notably in code completion, offering significant prospects to enhance developer efficiency~\cite{DBLP:journals/pacmpl/BarkeJP23, wechat-code-completion-survey}. These code-centric LMs (Code LMs)~\cite{chen2021evaluating,li2023starcoder} are predominantly trained using causal language modeling techniques, enabling them to predict subsequent code tokens by considering the context within the same file, referred to as the \textit{in-file context}.

Repository-level code completion~\cite{zhang-etal-2023-repocoder, ding2023crosscodeeval, liu2023repobench} extends beyond in-file context, aiming to synthesize unfinished code within the comprehensive context of the entire codebase. This involves leveraging the \textit{cross-file context}, which encapsulates high-level abstractions of various code constructs such as classes, functions, modules and similar code chunks within the other file. Recognizing that the functionality of a single line of code can be significantly influenced by this broader context is crucial. Consequently, the integration of \textit{cross-file context} is imperative for enhancing the precision and relevance of code completion systems.


Although \citet{zhang-etal-2023-repocoder,10.5555/3618408.3619722} have leveraged the Retrieval-Augmented-Generation (RAG) strategy to enhance code completion prediction accuracy through few-shot prompting\cite{10.5555/3495724.3495883}, it presents the challenge, called \textit{Context-Latency Conundrum}. This refers to the trade-off between richer context improving predictions and longer prompt length increasing inference times. A survey indicated that 85\% expect such tools to provide results within 200 milliseconds \cite{10.1145/3611643.3616280}, highlighting the importance of balancing context richness with inference latency.

\begin{figure*}[t]
	\centering
	\includegraphics[width=\linewidth]{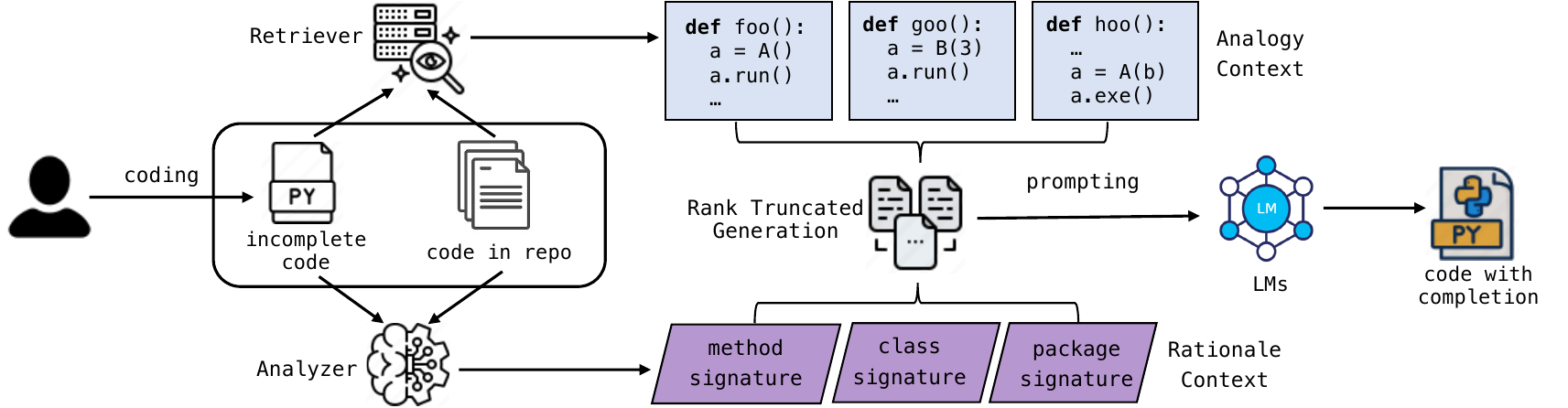}
 \vspace{-5mm}
	\caption{The workflow of \ToolName}
 \vspace{-2mm}
	\label{fig:workflow}
\end{figure*}

To address the challenge of the \emph{Context-Latency Conundrum}, we have devised a novel methodology that is \emph{dual context-aided}, proficiently capturing comprehensive contextual information within a constrained token space.
Our approach draws inspiration from the complex learning paradigms inherent to human cognition, which involve both the examination of analogous examples and the grasp of underlying concepts. 
The core of our innovation is the synthesis of two pivotal types of cross-file context:
the \emph{analogy context}, sourced from code similarity analysis~\cite{zakerinasrabadi2023systematic} that identify code segments with analogous functionality, and
the \emph{rationale context}, which provides a deep semantic understanding of class taxonomies and API interactions. 
Our procedure initiates with the aggregation of these critical contexts and incorporates few-shot prompting to educate the LM. However, the sheer volume of this information risks extending the prompt to an unwieldy size, potentially compromising efficiency. To mitigate this, \ToolName employs a strategic prompt formulation termed as \emph{rank truncated generation(RTG)}. This technique selectively filters the most pertinent context for the current task, thus condensing the data into a concise prompt designed to fit within a limited token window while ensuring high completion accuracy.

\ToolName's adeptness in critical context integration is validated by our benchmarks, where it achieves a 40.90\% to 59.75\% improvement in code exact match (EM) accuracy and realizes an enhancement in throughput efficiency of up to 24.4\% using CrossCodeEval benchmark. Notably, it attains this superior accuracy while utilizing only 75.4\% the context length typically required by other methods. This ensures contextually relevant prompt expansion and a significant 26.8\% increase in inference speed compared to traditional context type, allowing \ToolName for easier adoption in the fast response development scenarios.


In summary, the main contributions of this work are:
\begin{compactitem}
  \item We present a novel approach that fuses two types of cross-file context, \emph{analogy context} and \emph{rationale context}, to prompt LMs for code completion, which improves exact match accuracy with high efficiency.
   \item We introduce the \emph{rank truncated generation} for prompt construction, which underpins our model's ability to deliver high completion accuracy with concise prompts.
    \item In comparative evaluations, \ToolName surpasses existing methods by utilizing merely 75.4\% of the context length typically required, while simultaneously enhancing inference speed by 26.8\%. Importantly, \ToolName has been effectively implemented within a large organization, streamlining their daily software development workflows.
\end{compactitem}



\section{Preliminaries}
\label{label:preliminary}

To convenience the demonstration of our approach, we introduce several important concepts as preliminaries.
\begin{definition}(Completion Environment)
A completion environment $Env$ is $(s_e, F_r)$, where $s_e$ is the content of the file under the edit and $F_r$ is the list of source code files.
\end{definition}

Intuitively, the completion environment formalizes the scenario of writing the code in a given repository. 
Following existing studies on the code completion~\cite{zhang-etal-2023-repocoder}, we introduce the concepts of \emph{chunk} and \emph{chunk cover}.

\begin{definition}(Code Chunk)
\label{def:chunk}
    A code chunk is a code snippet containing a fixed line of code.
\end{definition}

\begin{definition}(Code Chunk Cover)
\label{def:cc}
    Given a completion environment $(s_e, F_r)$, a chunk size bound $\ell_{ck}$, and a sliding step $\eta$, all the source files are covered by a set of code chunks $\widehat{CCK}$. The subsequent portion of a specific chunk, denoted as $succ(ck)$, effectively illustrates the coding outcome when considering the code elements commonly shared between chunk $ck$ and its successor $succ(ck)$.
    
    For each $ck \in \widehat{CCK}$, we have (1) $ck$ has $\ell_{ck}$ lines of code, and (2) $ck$ has at most one successor code chunk that has $(\ell_{ck} - \eta)$ common lines of code with $ck$.
\end{definition}

As shown by \cref{def:chunk} and \cref{def:cc},
a chunk captures the local syntactic feature and shows the programming intent of the developers to some extents. The successor code chunk of a specific code chunk $ck$ actually demonstrates the coding result in the presence of the code commonly contained by $ck$ and $succ(ck)$. In our completion environment $(s_e, F_r)$, we denote the Code Chunks in $s_e$ that end at the completion point as unfinished code chunk $ck^{*}$, which reveals the code at the editing point. In a code completion task, our aim is to generate and append a token sequence to $ck^{*}$ according to the completion environment.

\section{Methodology}
\label{sec:method}

In this section, we present our technical design in detail. Our key idea stems from an observation of software development practices. As programmers begin contributing to a repository, they must demonstrate two essential skills.
Firstly, programmers must be proficient in the repository's architecture, encompassing cross-file modules, libraries, and APIs. A comprehensive understanding of this repository-level information is vital, as it allows them to write code accurately within the corresponding development environment, avoiding any misconceptions or errors—often referred to as ``hallucination" in programming contexts.
Secondly, programmers should become progressively acquainted with the repository. Drawing inspiration from similar modules within the repository, they can emulate previous designs and implementations when writing code for their specific tasks.


To align with the logical process of human programming, we present \ToolName, our repository-level code completion approach. This approach simultaneously leverages completed tokens and similar code in the repository to guide the completion process. We introduce two key concepts, namely \emph{rationale contexts} and \emph{analogy contexts}, which indicate the available program constructs and similar code snippets in the repository relevant to the current completion environment. These two contexts complement each other and provide valuable guidance for the completion process.

The workflow of \ToolName is depicted in \cref{fig:workflow}. It can be divided into three stages: rationale context analysis, analogy context retrieval, and rank truncated generation. We will provide detailed explanations of each stage.

\begin{figure*}[t]
	\centering
	\includegraphics[width=0.9\linewidth]{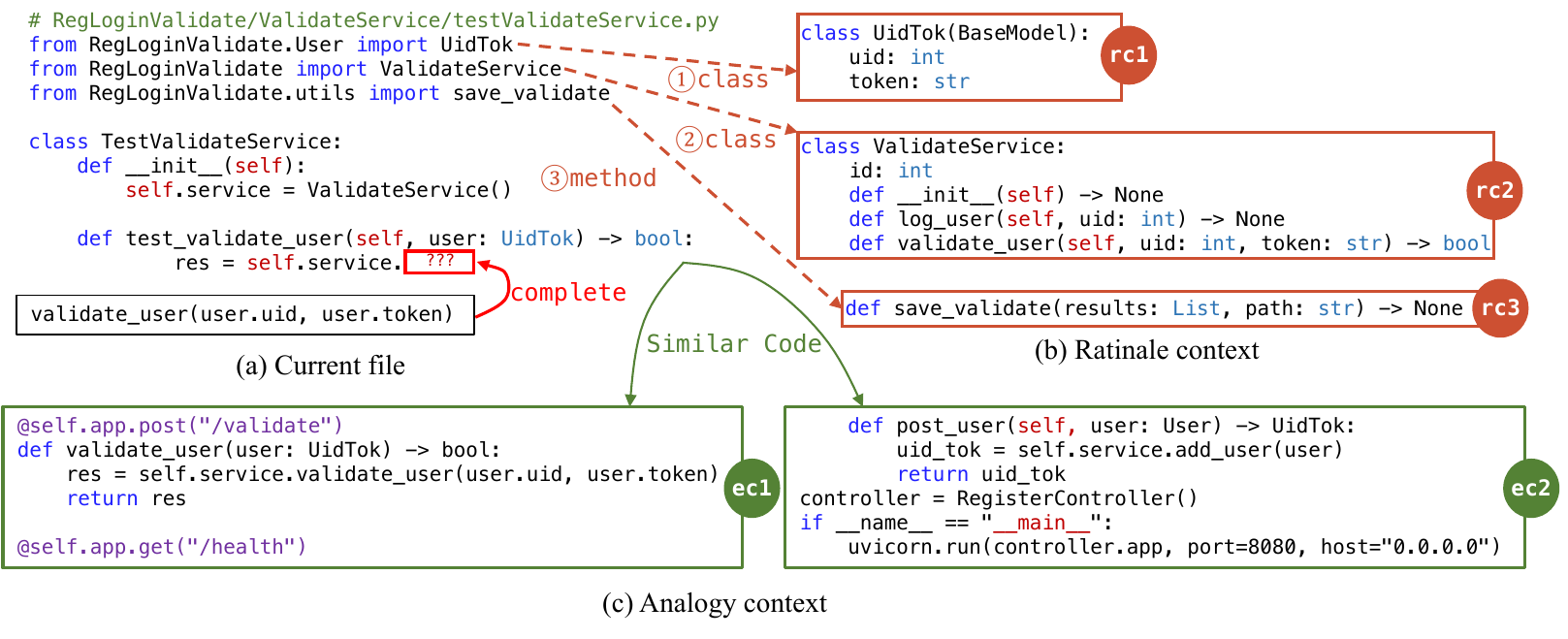}
 \vspace{-4mm}
	\caption{Illustrative Example of Rationale Context and Analogy Context in Use}
 \vspace{-2mm}
	\label{fig:example}
\end{figure*}
 
\subsection{Rationale Context Analysis}
\label{subsec:rca}

In software development, programmers need to be aware of the relevant program constructs accessible within their coding environment while editing a source file. This is established through import statements that bring in packages, modules, or header files, delineating the available classes and methods. We take the Python code completion as an example. By mimicking the process of analyzing import statements as humans, we can facilitate the code completion with the derived usable program constructs. 
Formally, we propose the concept of \emph{rationale context} to formalize such important ingredient for the code completion.


\begin{definition}(Rationale Context)
\label{def:rc}
Given a completion environment $Env:=(s_e, F_r)$,
a rationale context $\Gamma_r$ is a triple $(\Sigma_M, \Sigma_C, \Sigma_P)$, where $\Sigma_M$, $\Sigma_C$, and $\Sigma_P$ are the signatures of the methods, classes, and packages.
Particularly, the signature of a class is the concatenation of the signatures of the fields and methods offered by a class,
while the signature of a package is the concatenation of the signatures of the classes and methods offered by the package.
\end{definition}

Given a completion environment $(s_e, F_r)$, we compute the rationale context $\Gamma_r:=(\Sigma_M, \Sigma_C, \Sigma_P)$ by analyzing import statements in the edited file $s_e$.
According to \cref{def:rc}, the computation process is quite straightforward and can be achieved in a hierarchical manner.
For more detailed illustrations, we show a concrete example in \cref{fig:example}(a) and (b).
In the current file shown in \cref{fig:example}(a), we can derive three import statements, which import two classes and methods from other modules in the repository.
Such import statements actually determine the scope of the tokens that can be used at the current completion environment.
After localizing the definitions of the classes and methods,
we can extract their signatures and form a rationale context shown in \cref{fig:example}(b).
In this example, we obtain $\Gamma_r = (\{ rc3 \}, \ \{ rc1, rc2 \}, \ \emptyset)$.
Due to the space limit, we do not show the example of package signature with this example. We introduce the extraction of rationale context in \cref{subsec:implementation}.

\subsection{Analogy Context Retrieval}
\label{subsec:ecr}
When the programmers go through the repository,
they tend to concentrate on code snippets with a fixed size, i.e.,code chunks.
According to the unfinished code chunk $ck^{*}$ in the currently editing file,
we can discover several similar code chunks in other source files,
of which the successors offer informative guidance to the completion with their last $\eta$ lines of code.
Formally, we introduce the concept of the \emph{analogy context} as follows to formalize our intuition.
\begin{definition}(Analogy Context)
A analogy context $\Gamma_{a}$ is a set of code chunks, where a code chunk $ck \in \Gamma_{a}$ is highly similar to $ck^{*}$.
For a given similarity function $sim$ and a threshold $\varepsilon$, $\Gamma_{a}$ is defined as follows:
\begin{equation}
    \Gamma_{a}= \{ succ(ck)  \ | \ ck \in \widehat{CCK}, \ sim(ck, ck^{*}) < \varepsilon  \}
\end{equation}
\end{definition}


\cref{fig:example}(c) shows the analogy context retrieved based on the content of the current file shown in \cref{fig:example}(a).
Specifically, the two chunks in the analogy context, namely $ec1$ and $ec2$ share the similar tokens with the ones in the current file,
such as ``user'' and ``service''.
The high similarities imply that the chunks in the analogy context should be assigned with more attentions during the prompting process,
as they are likely to exhibit similar and even the same functionalities as the code in the current file.


\subsection{RTG-Empowered Completion}
\label{subsec:rtg}

As demonstrated by \cref{subsec:rca} and \cref{subsec:ecr}, the two kinds of contexts offer insightful clues to guide the completion for a specific completion environment.
However, code completion has a high requirement on the efficiency in real-world scenarios.
If we simply concatenate the rationale context with the analogy context in the prompt construction,
the prompt would be quite lengthy, and thus, introduce extra overhead in the LLM inference phase.

To utilize the two contexts efficiently,
we propose a prompt construction technique, named \emph{rank truncated generation} (RTG), to merge the two contexts into a prompt with a fixed size,
which can be formalized by the concept of \emph{truncated dual context}(TDC) as follows.
\begin{definition}(Truncated Dual Context)
\label{def:tdc}
    Given a rationale context $\Gamma_{r} = (\Sigma_M, \Sigma_C, \Sigma_P)$, an analogy context $\Gamma_{a}$, and the last chunk $ck^{*}$ at the end of the currently editing file, a truncated dual context is a set $\Gamma_{td}$ with the largest size satisfying the following constraints:
    \begin{itemize}
        \item (Truncation constraint) The total length of all the items in $\Gamma_{td}$ is bounded by the length bound $L$, i.e.,
        \begin{equation}
            \sum_{e \in \Gamma_{td}} len(e) \leq L
        \end{equation}
        Here the function $len$ measures the token number.
        \item (Rank constraint) For each item $e$ in $\Gamma_{td}$, all the items that are more relevant to $ck^{*}$ should belong to $\Gamma_{td}$. That is, for $e \in I$, we have
        \begin{equation}
            \forall e' \in U.\ r_{ck^{*}}(e') < r_{ck^{*}}(e) \rightarrow e' \in \Gamma_{td} 
        \end{equation}
        Here $U = \Sigma_M \cup \Sigma_C \cup \Sigma_P \cup \Gamma_{a}$. The function $r_{ck^{*}}$ is a scoring function induced by $ck^{*}$ that quantifies the relevance of a chunk with respect to $ck^{*}$.
    \end{itemize}
\end{definition}
Following \cref{def:tdc},
we instantiate the scoring function $r_{ck^{*}}$ by measuring the relevance between a code chunk and $ck^{*}$,
which can be implemented in various manners, 
such as choosing various distances and existing light-weighted embedding models.
In our evaluation, we evaluate the impact of different choices and discuss the results in detail.

In terms of the example in \cref{fig:example}, our RTG strategy will select $rc1$, $rc2$, and $ec1$ to create the truncated dual context.
Then, \ToolName follows the implementation in $ec1$ and generates a token sequence \texttt{validate\_user(user.uid, user.token)},
which invokes the method \texttt{validate\_user} upon \texttt{self.service} and passes \texttt{user.uid} and \texttt{user.token} as the arguments.
Notably, the method \texttt{validate\_user} and the fields \texttt{uid} and \texttt{token} of an \texttt{UidTok} are all offered in the rationale context.
With the guidance of the truncated dual context,
\ToolName{} can not only learn from the similar implementations but also tend to choose valid program constructs,
which can achieve better completion performance in the repository level.
For a more comprehensive understanding of the RTG setting and their implementation, the reader is referred to \cref{subsec:rtg_setting}.
\section{Experiments and Results}
\label{experiments}

This section presents the details of implementations and the main results of our evaluation.

\subsection{Implementation} 
\label{subsec:implementation}
We present methods to generate \emph{rationale} and \emph{analogy contexts}. The \emph{rationale context} evolves from our novel \GraphName, a specialized code property graph\cite{cpg} variant. Details on \GraphName\ are in the appendix and outlined in \cref{sec:graph}. It captures key interrelations within a repository's structure.
Precise locales in the graph are recorded, allowing us to trace each relationship's origin. For a file with a task hole, we extract relationships up to the task hole line to form the \emph{rationale context}. The \emph{analogy context} discussed in \cref{subsec:ecr} uses BM25 search results, detailed by \citet{ding2023crosscodeeval}. We provide an example of a generated prompt in \cref{appendix:prompt-example}, which corresponds to the code completion task example depicted in \cref{fig:example}.

The consideration of the performance impact associated with the generation of \GraphName\ is outside the scope of this experiment since the graph is pre-generated. The graph update time is inconsequential, at roughly 0.04 milliseconds. Nevertheless, in practical scenarios that necessitate frequent updates, the performance implications should be taken into account.

Having established the methodology for generating the \emph{rationale and analogy contexts}, we now turn our attention to assessing the performance of \ToolName{} on a state-of-the-art benchmark designed to rigorously test repository-level code completion capabilities.


\subsection{Benchmark}
\label{subsec:benchmark}
We assessed the efficacy of \ToolName using the SOTA CrossCodeEval\cite{ding2023crosscodeeval} benchmark. CrossCodeEval is a comprehensive dataset for evaluating code completion frameworks at the repository level. It includes Python, Java, TS, and C\# code snippets and focuses on the ability to understand cross-file context for accurate code predictions. The dataset presents real world programming challenges in multiple languages, requiring advanced comprehension of software repository interdependencies. By leveraging the Python subset of CrossCodeEval, which includes 471 repositories, 1,368 files, and 2,665 examples, our research seeks to conduct a thorough evaluation and substantiation of \ToolName's ability to generate accurate and context-sensitive code completions. Since our approach does not rely on any language-specific features, it can be easily generalized to other programming languages. We employ Code Match and Identifier Match as evaluation metrics, following the definitions provided in CrossCodeEval.

\subsection{Configuration}
\textbf{Model Selection:}
\label{subsec:model}
The repository collected in CrossCodeEval spans from March 5, 2023 to June 15, 2023. To prevent potential data leaks, Code LMs released before mid-2023 were selected. They are StarCoderBase with 1B, 3B and 7B size~\cite{li2023starcoder}, DeepSeek-Coder with 1B and 7B size~\cite{deepseek-coder}, CodeLlama with 7B~\cite{rozière2023code}.  The evaluation focused on models that support 8k input tokens to incorporate information from all cross files. 


    

\textbf{RTG Setting:} 
\label{subsec:rtg_setting}
As discussed in \cref{subsec:rtg}, for rationale context, we evaluate truncation size varies from 256 to 4096. We represent the context length distribution of the evaluated data in \cref{fig:distribution}. In the context of a 4096-token length, the majority of samples remain untruncated, allowing us to approximate it as ``Unlimited''. This implies that all information from context is retained without omission. We have set the maximum sequence length for infile context to 2048 tokens, thereby preventing any truncation of infile information that could affect the results.

To evaluate the RTG score function $r_{ck^{*}}$, we consider four distinct Score Functions as follows:
\begin{compactitem}
  \item \textbf{Oracle}: In this idealized scenario, every context $e \in U$ is used as a prompt for the language model, and the resulting code's Edit Similarity metric with the Ground Truth is taken as its score. The Oracle represents the theoretical best-case performance for any Score Function but is impractical for real-world use due to significant computational demands.
  \item \textbf{Semantic Similarity}: We encode the semantic representation of each context $e$ and the unfinished code chunk $ck^*$ using an encoding model. The cosine similarity of these embeddings serves as the score. We employ encoding models such as CodeBERT~\cite{feng-etal-2020-codebert} and UniXcoder~\cite{guo-etal-2022-unixcoder}.
  \item \textbf{Lexical Similarity}: This score is determined using simple textual comparisons, particularly Jaccard Similarity and Edit Similarity measures.
  \item \textbf{Random}: Scores are assigned randomly, serving as a baseline to underscore the effectiveness of the other Score Functions.
\end{compactitem}

\begin{table*}[ht]
\centering
\caption{Performance with Analogy Context, Rationale Context and Dual Context on Benchmark over All Code LMs. }
    \resizebox{0.88\linewidth}{!} {
\begin{threeparttable}
\begin{tabular}{lc|cc|cccc|ccc}
\toprule
\multicolumn{1}{c}{\textbf{Model}} & 
\multicolumn{1}{c}{\textbf{Context}} &
\multicolumn{2}{c}{\textbf{Code Match}} & 
\multicolumn{1}{c}{} &
\multicolumn{2}{c}{\textbf{Identifier Match}} &
\multicolumn{1}{c}{} &
\multicolumn{3}{c}{\textbf{No. of CTXs}} \\
\multicolumn{1}{c}{} &
\multicolumn{1}{c}{} &
\multicolumn{1}{c}{\textbf{EM}} & 
\multicolumn{1}{c}{\textbf{ES}} & 
\multicolumn{1}{c}{\textbf{EM}} & 
\multicolumn{1}{c}{\textbf{P}}& 
\multicolumn{1}{c}{\textbf{R}}& 
\multicolumn{1}{c}{\textbf{F1}} &
\multicolumn{1}{c}{\textbf{AC}}&
\multicolumn{1}{c}{\textbf{RC}}\\
\midrule
\multirow{3}{*}{StarCoder-1B}   
  & AC & 11.18 & 62.46 & 20.04 & 54.82 & 51.74 & 51.55 & 4.85 & 0 \\
  & RC & 10.54 & 60.40 & 19.25 & 53.27 & 50.12 & 49.73 & 0 & 4.32 \\
  & DC & 17.86 (\textcolor{red}{+6.68}, \textcolor{blue}{+7.32}) & 66.21 & 27.65 & 60.80 & 58.19 & 57.57 & 4.81 & 4.25 \\ \hline
\multirow{3}{*}{StarCoder-3B} 
  & AC & 14.26 & 65.11 & 23.49 & 57.77 & 55.37 & 54.88 & 4.85 & 0 \\
  & RC & 14.71 & 63.41 & 23.30 & 57.50 & 54.18 & 53.96 & 0 & 4.32 \\
  & DC & 21.39 (\textcolor{red}{+7.13}, \textcolor{blue}{+6.68}) & 69.29 & 31.78 & 64.47 & 61.91 & 61.41 & 4.81 & 4.25 \\ \hline
\multirow{3}{*}{StarCoder-7B} 
  & AC & 16.02 & 66.64 & 25.7 & 59.80 & 56.92 & 56.73 & 4.85 & 0  \\
  & RC & 17.37 & 65.20 & 25.93 & 59.90 & 56.78 & 56.44 & 0 & 4.32  \\
  & DC & 24.20 (\textcolor{red}{+8.18}, \textcolor{blue}{+6.83}) & 70.82 & 34.60 & 66.67 & 63.98 & 63.55 & 4.81 & 4.25 \\ \hline
\multirow{3}{*}{DeepSeek-1B}   
  & AC & 14.82 & 65.01 & 23.90 & 58.11 & 54.89 & 54.85 & 4.85 & 0 \\
  & RC & 16.14 & 63.86 & 24.73 & 58.32 & 55.39 & 55.01 & 0 & 4.30 \\
  & DC & 22.59 (\textcolor{red}{+7.77}, \textcolor{blue}{+6.45}) & 68.85 & 32.68 & 64.29 & 61.91 & 61.36 & 4.80 & 4.20 \\ \hline
\multirow{3}{*}{DeepSeek-7B}    
  & AC & 19.70 & 69.11 & 29.57 & 62.99 & 59.55 & 59.74 & 4.85 & 0 \\
  & RC & 21.13 & 68.09 & 30.32 & 63.13 & 59.86 & 59.74 & 0 & 4.30 \\
  & DC & 27.92 (\textcolor{red}{+8.22}, \textcolor{blue}{+6.79}) & 73.09 & 39.21 & 69.42 & 66.50 & 66.39 & 4.80 & 4.20  \\ \hline
 \multirow{3}{*}{CodeLlama-7B}   
  & AC & 17.60 & 67.23 & 27.32 & 60.28 & 57.47 & 57.40 & 4.85 & 0  \\
  & RC & 18.31 & 65.60 & 26.64 & 59.46 & 56.32 & 56.17 & 0 & 4.30  \\
  & DC & 24.80 (\textcolor{red}{+7.2}, \textcolor{blue}{+6.49}) & 71.05 & 34.93 & 66.48 & 64.01 & 61.61 & 4.80 & 4.20  \\
\bottomrule
\end{tabular}
\begin{tablenotes}
\item \textcolor{red}{+} Comparison with AC, \textcolor{blue}{+} Comparison with RC, CTX: average selected candidates in different contexts.
\end{tablenotes}
\end{threeparttable}
}
\vspace{-3mm}
\label{tab:benchmark_performance}
\end{table*}

\subsection{Results and Analysis}
We present our experimental results and add detailed illustrations on the evaluation findings.

\subsubsection{Performance}
\textbf{Effectiveness of Dual Context:} 
We evaluated the performance of various models with three distinct context settings under a truncation size of 4096: Analogy Context (\textbf{AC}), where only analogy context are used for code completion; Rationale Context (\textbf{RC}), where only rationale contexts are used; and Dual Context (\textbf{DC}), which combines both analogy and rationale contexts. The first two settings served as baselines in our study. As discussed in \cref{subsec:rtg_setting}, the minimal information loss at this truncation size allows for an effective comparison of how different contexts influence the outcomes.

\begin{figure*}[t]
	\centering
	\includegraphics[width=0.9\linewidth]{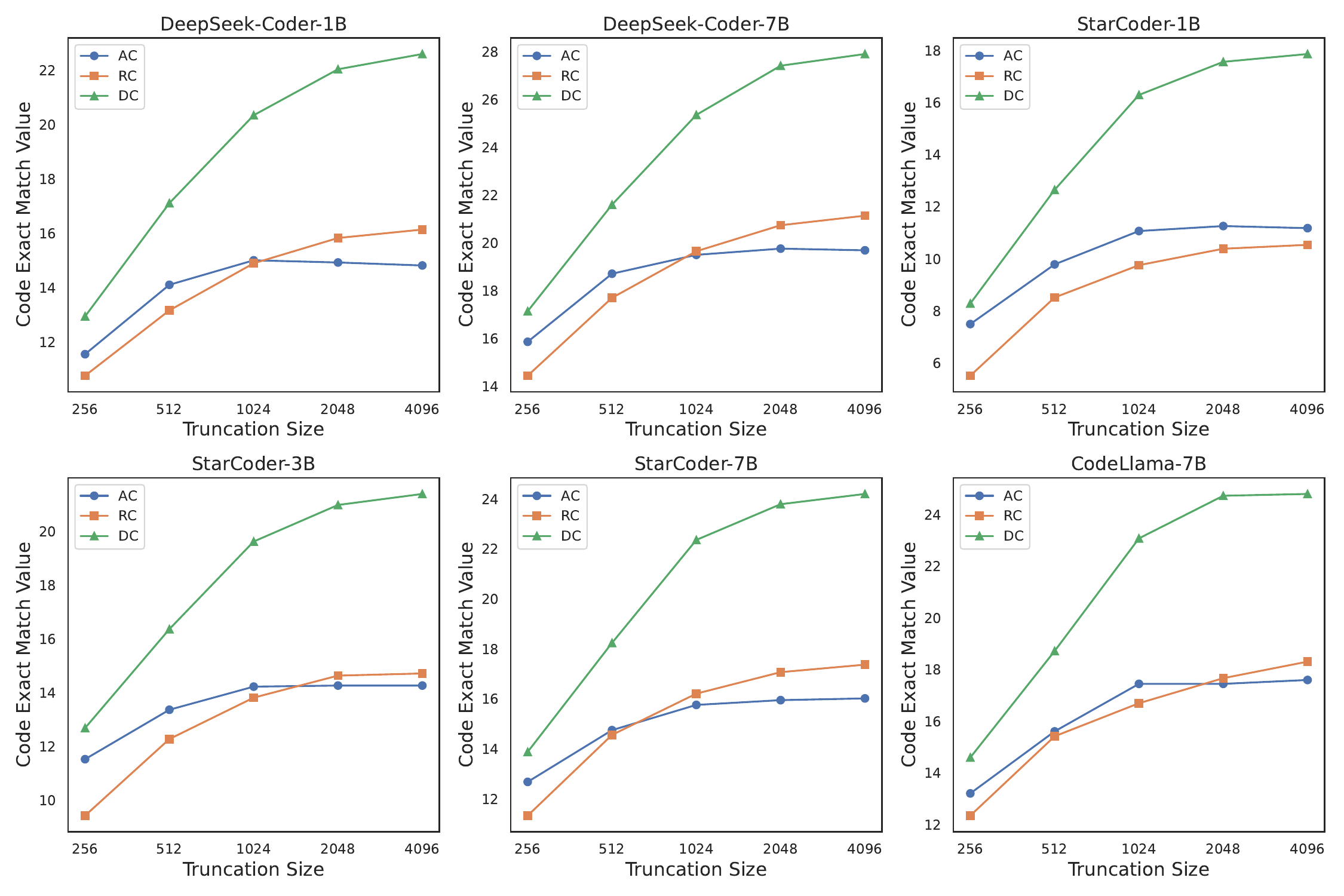}
 \vspace{-2mm}
	\caption{The comparison of Code\_EM performance on all truncation sizes for AC, RC and DC.}
 \vspace{-2mm}
	\label{fig:comparison_of_all_sizes}
\end{figure*}

\begin{figure}[t]
	\centering
	\includegraphics[width=0.65\columnwidth]{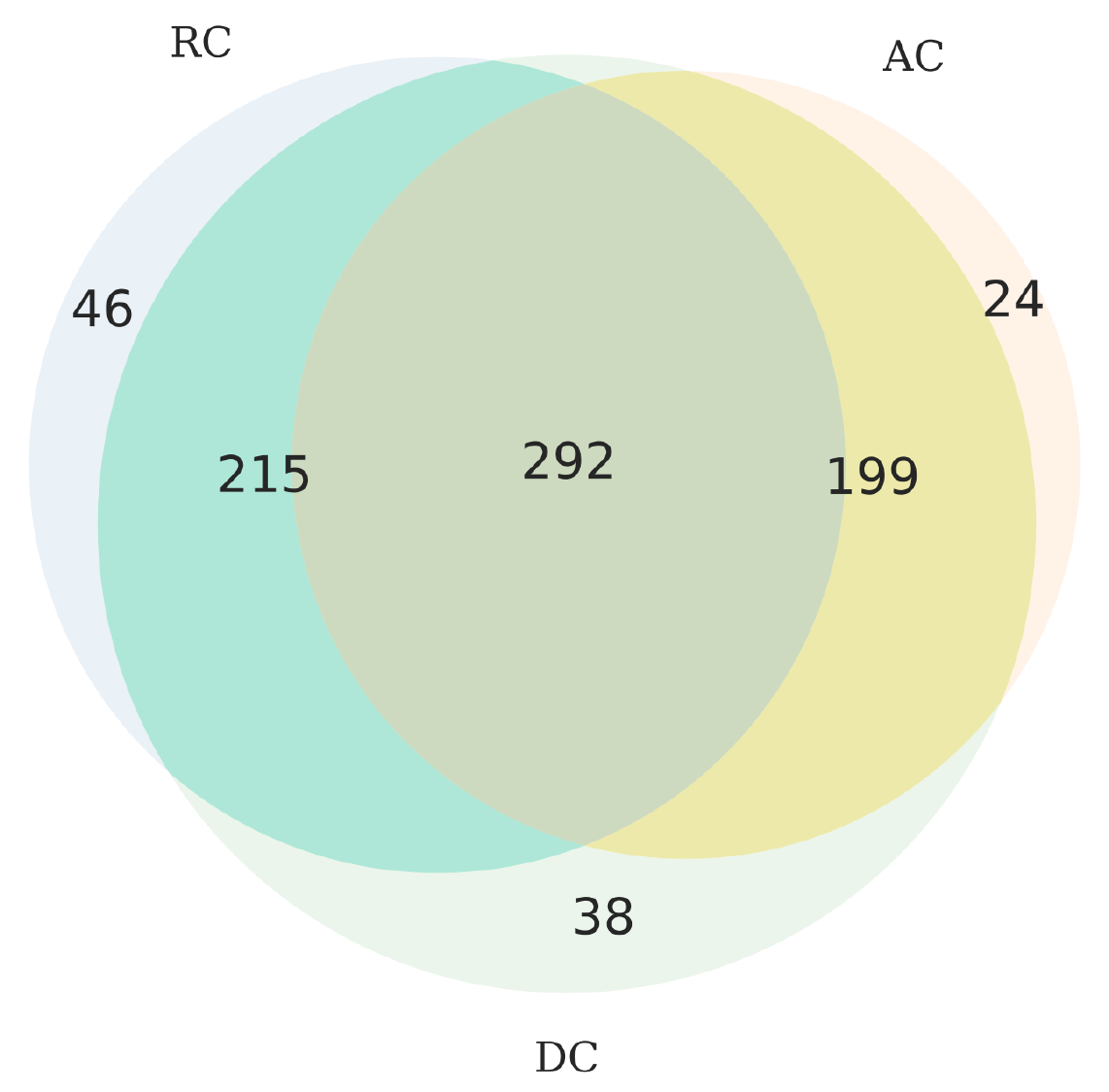}
  \vspace{-2mm}
	\caption{Convergence Analysis of Code\_EM for Generated Cases by AC, RC and DC on DeepSeek-Coder-7B.}
	\label{fig:convergence}
  \vspace{-7mm}
\end{figure}

\begin{figure}[t]
	\centering
	\includegraphics[width=0.95\columnwidth]{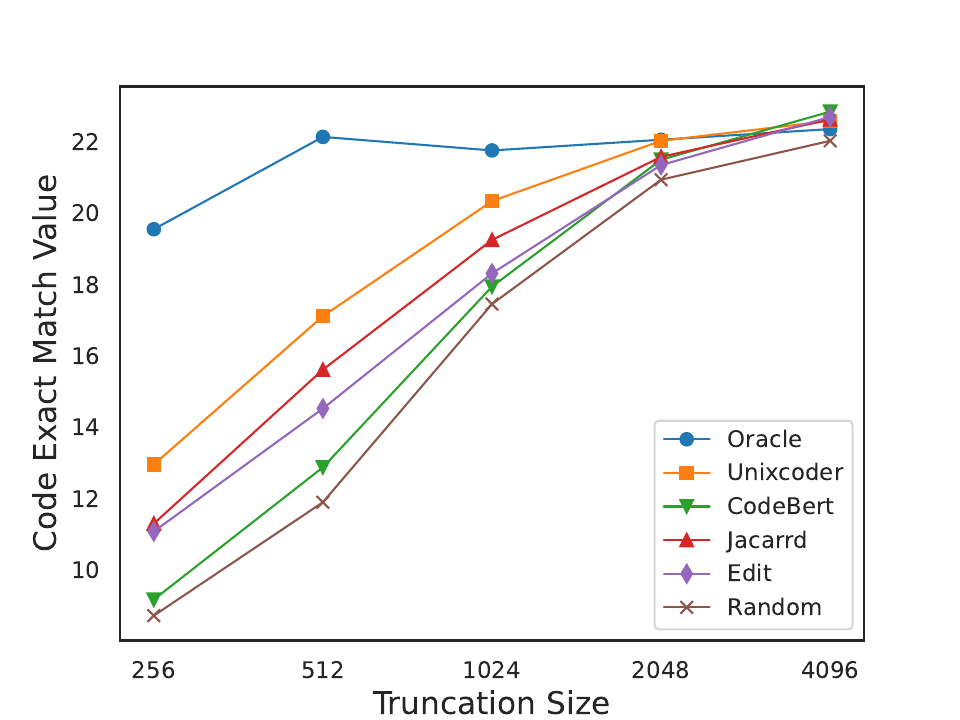}
  \vspace{-2mm}
	\caption{Comparative Assessment of Code\_EM Values Across Various Truncation Sizes Using DC with Diverse Scoring Functions.}
	\label{fig:ranking}
  \vspace{-2mm}
\end{figure}

\begin{figure}[t]
\centering
\vspace{-2mm}
\includegraphics[width=0.85\columnwidth]{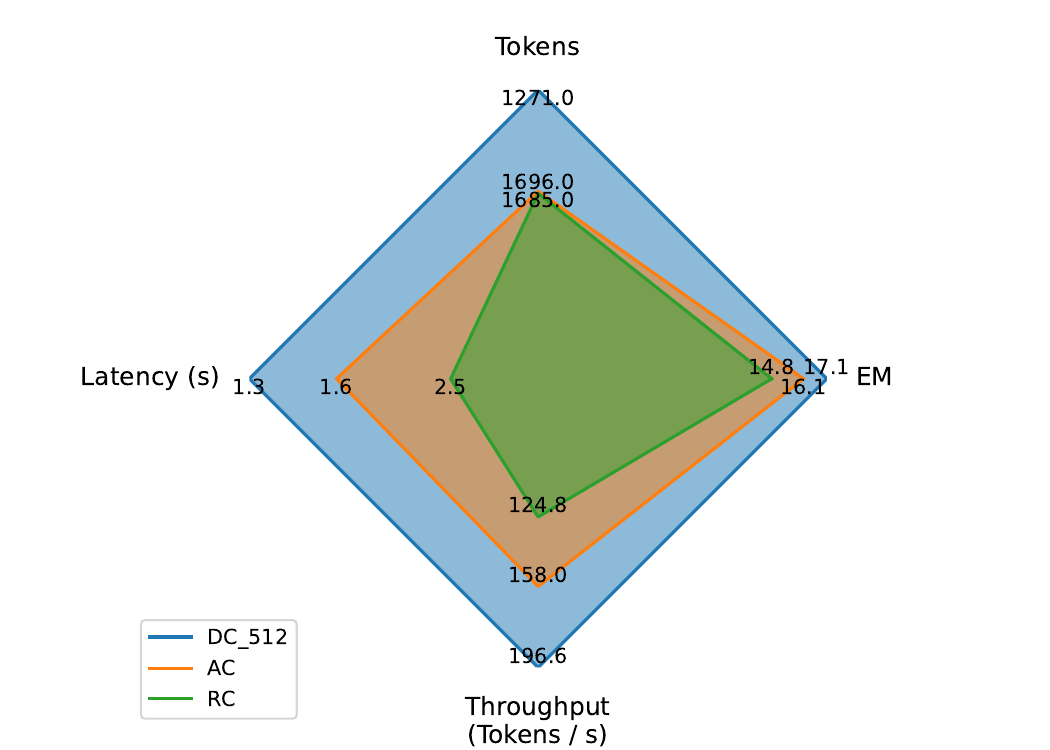}
\caption{Inference Efficiency of DeepSeek-Coder-1B on different RAG Strategies with Batch\_size = 8. }
\label{fig:radarchart}
\end{figure}

\begin{figure}[t]
\centering
\includegraphics[width=0.7\columnwidth]{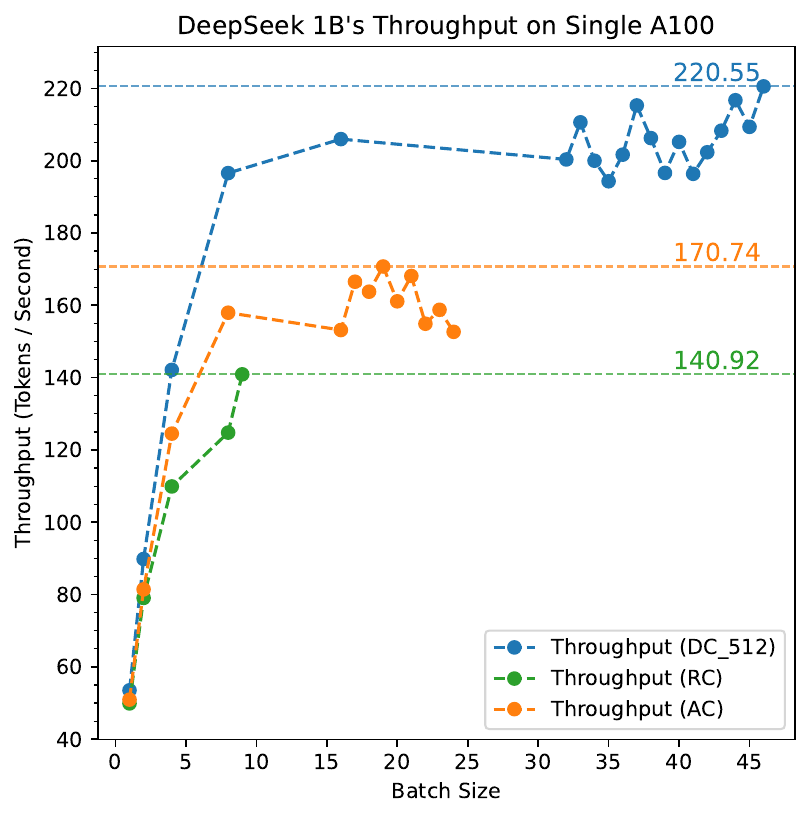}
\vspace{-3mm}
\caption{Throughput efficiency of DeepSeek-Coder-1B with RAG strategies on NVIDIA A100 GPU, truncated due to OOM.}
\vspace{-5mm}
\label{fig:throughput_1b_a100_no_ap}
\end{figure}

As shown in \cref{tab:benchmark_performance}, Dual Context significantly outperforms baseline contexts AC or RC across various Code LMs. This improvement suggests that analogy and rationale information are complementary; their combination leads to superior model performance. Specifically, for the Code Exact Match (Code\_EM) metric, Dual Context registers a substantial increase, rising from 6.68 to 8.22 over the analogy context and from 6.45 to 7.32 over the rationale context on different Code LMs. This equates to a percentage enhancement from 40.90\% to 59.75\% over the analogy context and from 32.13\% to 69.45\% over the rationale context. The complete experimental results for the three context settings on the DeepSeek-Coder-7B model with all truncation sizes are detailed in \cref{appendix-baseline}.


\textbf{Scalability Across Truncation Sizes:} We assess the performance scalability of AC, RC and DC across different truncation sizes, from 256 to 4096 tokens, for various Code LMs detailed in \cref{subsec:model}. The results, as depicted in \cref{fig:comparison_of_all_sizes}, show that DC's performance consistently outstrips that of AC and RC with increasing truncation size. \textit{Notably, DC at a truncation size of 512 tokens outperforms both AC and RC at 4096 tokens, demonstrating the efficiency of Dual Context in encapsulating vital information from AC and RC in a more compact form}. This efficiency in truncation size not only boosts performance but also inference speed, which we elaborate on in \cref{section:inference_efficiency}. Furthermore, the DeepSeek-Coder model outshines other LMs of comparable size, implying that integrating cross-file data during pretraining significantly enhances performance on repository-level code completion tasks.


\textbf{Unique Contributions of Dual Context:} The convergence trends for Code EM instances within the DeepSeek-7B model are illustrated in \cref{fig:convergence}. Analyses of the three context types—AC, RC, and DC—reveal that RC accounts for 553 exact matches, AC for 515 and DC stands out with 744. The superiority of DC is attributed to two primary factors. First, by amalgamating contextual information from both AC and RC, DC provides a richer, complementary context that allows Code LMs to prioritize the most pertinent information for diverse sample types. Second, DC enables the resolution of certain cases that cannot be adequately addressed using exclusively AC or RC. In \cref{fig:convergence}, 38 such cases are evident, where the combined insights from both contexts are essential for precise code completion. For convergence results related to other Code LMs, readers are directed to the \cref{appendix:convergence}.
 


\subsubsection{Inference Efficiency}
\label{section:inference_efficiency}
\textbf{Inference Efficiency with Truncated Contexts:} DC at a truncation size of 4096 tokens has the best accuracy but demands substantial computational resources. To align with the limited computational resources typically available, we reduced the truncation size to 512 tokens and assessed the DeepSeek-Coder-1B model's performance. As shown in \cref{fig:radarchart}, Truncated DC (DC\_512) surpasses AC and RC in EM performance, tokens length, inference throughput, and latency. The Radar Chart confirms that \ToolName achieves not only higher accuracy but also an increased throughput by 24.4\% over AC and 57.5\% over RC, as well as a reduced latency with a speed optimization of 26.8\% against AC and 95.3\% against RC, with a batch size of 8.



\textbf{Throughput-Oriented Efficiency:} In scenarios with high user concurrency, throughput performance is paramount. The DeepSeek-Coder-1B model on a single A100 GPU demonstrates this, where Truncated DC (DC\_512) achieves throughput increases of 29.2\% and 56.5\% over AC and RC, respectively, as evidenced in \cref{fig:throughput_1b_a100_no_ap}.
DC\_512 not only tolerates larger batch sizes—up to 2$\times$ and 5$\times$ larger than AC and RC, which is critical for dealing with traffic spikes—but also demonstrates this capability under memory constraints, as illustrated by the OOM-truncated lines in \cref{fig:throughput_1b_a100_no_ap}. This confirms the efficiency of our truncation strategy in high-demand situations. Comprehensive performance metrics are presented in \cref{sec:effi}.



\subsubsection{Comparison of RTG Score Functions}
We evaluate score functions crucial for RTG's context retention decisions within truncation limits in \cref{sec:method}. \cref{fig:ranking} reveals that the Oracle function excels at a 512 truncation size, with no significant performance differences among strategies at 2048. This suggests that a refined score function is vital for selecting pertinent context, optimizing prompt length, and improving inference efficiency for Code LMs.
UniXcoder outshines other strategies (except Oracle) and Random ranks as the least effective, corroborating \cite{liu2023repobench}'s insights. UniXcoder's multi-modal pre-training, incorporating AST and comments, aligns well with RC, enhancing comprehension in Code LMs.
The strategies are effective in AC or RC exclusive scenarios, detailed in \cref{appendix:oracle-function}. Code LMs' indifference to information order implies a potentially lower impact of ordering strategies, with further analysis in \cref{appendix:ordering}.

\section{Related Work}
\label{related_work}

\textbf{Code Language Models:} 
In recent years, Code LMs have found extensive application in the field of software engineering, notably contributing to the enhancement of development efficiency, with code completion scenarios being particularly prominent~\cite{wechat-code-completion-survey}. A diverse range of Pretrained Code LMs has been deployed in this context (Codex\cite{chen2021evaluating}, StarCoder\cite{li2023starcoder}, CodeGen\cite{nijkamp2023codegen2}, CodeLlama\cite{rozière2023code}, In-Coder\cite{DBLP:conf/iclr/FriedAL0WSZYZL23}, SantaCoder\cite{DBLP:journals/corr/abs-2301-03988}, CodeGeex\cite{codegeex}, CodeFuse\cite{codefuse13b}, PanguCoder\cite{pangucoder}, DeepSeek-Coder\cite{deepseek-coder}). Many of these works have harnessed the Transformer Decoder architecture for training in the Next Token Prediction task using source code data, allowing them to directly apply their models to Code Completion tasks. Additionally, certain studies have incorporated the Fill-in-the-Middle task into their training stage (\cite{li2023starcoder}, \cite{rozière2023code}, \cite{deepseek-coder}, \cite{fim}) to better align with real-world code completion scenarios. 
Most of these Code LMs mainly concentrate on local information within individual files and lack comprehensive support for entire code repositories except DeepSeek-Coder \cite{deepseek-coder}, which make the first attempt to incorporate repository-level data construction in pre-training stage.


\textbf{Repository-level Context-aware Code Completion:}
Research into repository-level analysis has gained traction as a method for understanding code at scale(\citet{10.5555/2487085.2487127}, \cite{bairi2023codeplan}, \cite{gupta2023grace}, \cite{10.1145/3377811.3380342}). In the context of repository-level code completion, there are many works aim to utilize the vast amount of code available in current repositories to aid Code LMs in generating better completions(\cite{liu2023repobench}, \cite{DBLP:conf/aaai/PeiZLZK23}, \cite{ding2023crosscodeeval}). In the work by \cite{zhang-etal-2023-repocoder}, the authors propose an iterative retrieval generation pipeline to demonstrate high quality prompt for code completion. In \cite{10.5555/3618408.3619722}, example-specific prompts are generated by a learned Prompt Proposal Generator, introducing repository-level knowledge. Furthermore, in \cite{shrivastava2023repofusion}, they employ a fine-tuned Fusion-in-Decoder model to integrate information derived from the repository. Additionally, \cite{Ding2022CoCoMICCC} starts with the utilization of a Cross-file Context Finder to retrieve cross-file context based on the Project Context Graph. Subsequently, multiple cross-file pieces of information are compressed into embedding representations. These representations are then employed by a fine-tuned Code LM as k-v cache to enhance the code generation process. There are also other works that explore alternative sources to utilize useful information. \cite{zan2023privatelibraryoriented} introduce a private-library-oriented code completion scenario which generate prompt from private API Documentation. \cite{lu-etal-2022-reacc} retrieve similar code snippet from self-collected Source Code Database. As fast response development scenarios require efficient approaches, many of these existing methods may not be applicable, highlighting the growing need for more efficient approaches.


\section{Conclusion and Future Work}
\label{conclusion}


This paper presents \ToolName, a novel repository-level code completion approach. By extracting dual contexts from the completion environment and utilizing a new prompt generation strategy, \ToolName achieves high accuracy and efficiency. Evaluation results demonstrate its superior performance compared to state-of-the-art techniques. \ToolName has been integrated into a commercial company's production line and will soon be released as open source.

In the future, \ToolName can be extended from the following perspectives. 
First, a lightweight program analyzer can be employed to remove unnecessary program constructs from the rationale context, improving completion accuracy and efficiency. By identifying and eliminating constructs that cannot be invoked without specific object types, the completion process becomes more precise and faster.
Second, the suffix content beyond the unfinished code has not been utilized in this work, which is known as FIM(Fill-in-the-middle) \cite{fim} task. the relationship between crossfile, prefix, and suffix, as well as the integration of this information with Code LMs trained on FIM objectives, warrants further discussion.
Third, due to memory constraints and specific usage scenarios, the evaluation of \ToolName was limited to smaller Code LMs. However, it is expected that \ToolName utilizing larger Code LMs will achieve improved code completion performance. Further practical implementation are necessary to validate this assumption. \lming{pls add fim discussion here}


\section*{Broader Impact}

Our research advances code completion by integrating cross-file context, enhancing the utility of code generation tools without directly training language models. Consequently, our work bypasses the substantial computational resources typically associated with language model pretraining, thereby mitigating related environmental impacts.
While we acknowledge the risks associated with code generation, such as bias or insecurity, we have implemented measures to minimize these risks. Our approach primarily focuses on refining prediction mechanisms in existing models, rather than contributing to these risks at a comparable level to model training.

Our approach, which focuses on the precision of code completion through \textit{cross-file context}, enhances developer productivity without the negative broader impacts of increased energy consumption or greenhouse gas emissions. We believe our work sets a precedent for responsible AI development, balancing innovation with ethical considerations.


\bibliography{references}
\bibliographystyle{icml2024}

\newpage
\appendix
\onecolumn

\section{Experiment results of \ToolName{} over DeepSeek-Coder-7B on All Truncation Sizes}
\label{appendix-baseline}

\begin{table}[ht]
\centering
\footnotesize
\begin{tabular}{lcc|cc|cccc|cc}

\toprule
\multicolumn{1}{c}{\textbf{Size}} & 
\multicolumn{1}{c}{\textbf{Ranking}} & 
\multicolumn{1}{c}{\textbf{Context}} &
\multicolumn{2}{c}{\textbf{Code}} & 
\multicolumn{1}{c}{} &
\multicolumn{2}{c}{\textbf{Identifier}} &
\multicolumn{1}{c}{} &
\multicolumn{2}{c}{\textbf{No. of CTXs}} \\
\multicolumn{1}{c}{} &
\multicolumn{1}{c}{} &
\multicolumn{1}{c}{} &
\multicolumn{1}{c}{\textbf{EM}} & 
\multicolumn{1}{c}{\textbf{ES}} & 
\multicolumn{1}{c}{\textbf{EM}} & 
\multicolumn{1}{c}{\textbf{P}}& 
\multicolumn{1}{c}{\textbf{R}}& 
\multicolumn{1}{c}{\textbf{F1}} &
\multicolumn{1}{c}{\textbf{AC}}&
\multicolumn{1}{c}{\textbf{RC}}\\
\midrule
\multirow{15}{*}{256} & \multirow{3}{*}{CodeBERT}   
  & AC & 12.91 & 64.14 & 20.83 & 56.26 & 52.53 & 52.83 & 1.23 & 0\\
  & & RC & 13.02 & 63.55 & 20.68 & 55.96 & 51.95 & 52.32 & 0 & 1.32 \\
  & & DC & 13.62 & 64.34 & 21.35 & 56.52 & 52.74 & 53.05 & 0.9 & 0.66\\ \cline{2-11}
  & \multirow{3}{*}{Edit}
    & AC & 14.86 & 65.29 & 22.93 & 57.97 & 53.81 & 54.34 & 1.19 & 0 \\
  & & RC & 13.17 & 63.68 & 20.86 & 56.28 & 52.37 & 52.68 & 0 & 1.42 \\
  & & DC & 14.93 & 65.57 & 23 & 58.29 & 54.37 & 54.75 & 0.94 & 0.57 \\ \cline{2-11}
    & \multirow{3}{*}{Jacarrd}
    & AC & 14.63 & 65.24 & 22.89 & 57.92 & 53.93 & 54.31 & 1.11 & 0 \\
  & & RC & 13.58 & 64 & 21.46 & 56.67 & 52.67 & 53.06 & 0 & 1.2\\
  & & DC & 15.12 & 65.41 & 23.08 & 58.32 & 54.34 & 54.68 & 0.9 & 0.38 \\ \cline{2-11}
    & \multirow{3}{*}{Random}
    & AC & 12.46 & 63.81 & 20.49 & 55.66 & 51.77 & 52.13 & 1.06 & 0 \\
  & & RC & 12.8 & 63.4 & 20.3 & 55.99 & 51.77 & 52.2 & 0 & 1.01 \\
  & & DC & 12.12 & 63.8 & 19.96 & 56.07 & 52.04 & 52.49 & 0.69 & 0.44 \\ \cline{2-11}
    & \multirow{3}{*}{UniXcoder}
    & AC & 15.87 & 65.73 & 24.09 & 58.67 & 54.67 & 55.11 & 1.05 & 0 \\
  & & RC & 14.45 & 64.27 & 22.06 & 56.96 & 52.94 & 53.31 & 0 & 1 \\
  & & DC & 17.15 & 66.42 & 24.99 & 59.67 & 55.73 & 56.08 & 0.78 & 0.36\\ 
  \hline
\multirow{15}{*}{512} & \multirow{3}{*}{CodeBERT}   
  & AC & 16.81 & 66.56 & 25.59 & 59.77 & 56.43 & 56.55 & 2.79 & 0\\
  & & RC & 16.02 & 65.39 & 24.32 & 59.05 & 55.32 & 55.44 & 0 & 2.27 \\
  & & DC & 17.9 & 67.26 & 26.9 & 60.89 & 57.57 & 57.57 & 2.09 & 1.32 \\ \cline{2-11}
  & \multirow{3}{*}{Edit}
    & AC & 18.2 & 67.73 & 27.35 & 61.16 & 57.24 & 57.67 & 2.76 & 0\\
  & & RC & 16.14 & 65.31 & 24.17 & 58.78 & 55 & 55.17 & 0 & 2.39\\
  & & DC & 19.25 & 68.2 & 28.22 & 62.1 & 58.09 & 58.49 & 2.23 & 1.1 \\ \cline{2-11}
    & \multirow{3}{*}{Jacarrd}
    & AC & 18.39 & 67.76 & 27.73 & 61.28 & 57.68 & 57.95 & 2.65 & 0 \\
  & & RC & 16.47 & 65.34 & 24.69 & 59.18 & 55.46 & 55.62 & 0 & 2.09 \\
  & & DC & 20.23 & 68.76 & 29.98 & 62.91 & 59.38 & 59.56 & 2.15 & 0.81 \\ \cline{2-11}
    & \multirow{3}{*}{Random}
    & AC & 15.87 & 66.51 & 25.07 & 59.45 & 55.52 & 55.94 & 2.6 & 0 \\
  & & RC & 15.5 & 64.99 & 23.79 & 58.38 & 54.51 & 54.77 & 0 & 1.85 \\
  & & DC & 17.56 & 66.5 & 26.19 & 59.9 & 56.25 & 56.49 & 1.69 & 0.99\\ \cline{2-11}
    & \multirow{3}{*}{UniXcoder}
    & AC & 18.72 & 68 & 27.95 & 61.75 & 58.32 & 58.43 & 2.59 & 0 \\
  & & RC & 17.71 & 66.06 & 26.19 & 59.88 & 56.29 & 56.38 & 0 & 1.89 \\
  & & DC & 21.61 & 69.47 & 31.07 & 63.94 & 60.52 & 60.61 & 1.89 & 0.78\\ 
  \hline
\multirow{15}{*}{1024} & \multirow{3}{*}{CodeBERT}   
  & AC & 19.92 & 69.18 & 29.61 & 63.07 & 59.66 & 59.84 & 4.69 & 0\\
  & & RC & 18.42 & 66.46 & 27.2 & 60.84 & 57.49 & 57.44 & 0 & 3.25 \\
  & & DC & 23.26 & 70.78 & 33.4 & 65.72 & 62.6 & 62.51 & 3.83 & 2.37 \\ \cline{2-11}
  & \multirow{3}{*}{Edit}
    & AC & 19.77 & 69.07 & 29.79 & 63.22 & 59.71 & 59.95 & 4.69 & 0 \\
  & & RC & 18.54 & 66.63 & 27.28 & 61.24 & 57.71 & 57.73 & 0 & 3.39 \\
  & & DC & 23.49 & 70.98 & 33.81 & 66.06 & 62.6 & 62.69 & 4.09 & 2.22 \\ \cline{2-11}
    & \multirow{3}{*}{Jacarrd}
    & AC & 19.62 & 69.22 & 29.68 & 63.14 & 59.65 & 59.86 & 4.68 & 0 \\
  & & RC & 19.4 & 66.9 & 28.22 & 61.56 & 58.03 & 58.09 & 0 & 3.12 \\
  & & DC & 24.09 & 71.54 & 34.45 & 66.63 & 63.14 & 63.3 & 3.97 & 1.86 \\ \cline{2-11}
    & \multirow{3}{*}{Random}
    & AC & 19.81 & 69.14 & 29.87 & 63.19 & 59.74 & 59.96 & 4.68 & 0 \\
  & & RC & 18.99 & 66.76 & 27.47 & 61.28 & 57.77 & 57.81 & 0 & 2.87 \\
  & & DC & 22.66 & 69.89 & 32.87 & 65.3 & 61.71 & 61.85 & 3.29 & 2.08 \\ \cline{2-11}
    & \multirow{3}{*}{UniXcoder}
    & AC & 19.51 & 68.98 & 29.42 & 62.94 & 59.4 & 59.63 & 4.67 & 0 \\
  & & RC & 19.66 & 67.03 & 28.33 & 61.89 & 58.4 & 58.38 & 0 & 2.9 \\
  & & DC & 25.37 & 71.77 & 36.1 & 67.5 & 64.23 & 64.25 & 3.69 & 1.77\\ 
\bottomrule
\end{tabular}
\caption{Performance with Analogy Context, Rationale Context and Dual Context on Benchmark over DeepSeek-Code-7B.}
\label{tab:allExperiment1}
\end{table}

\begin{table}[ht]
\centering
\footnotesize
\begin{tabular}{lcc|cc|cccc|cc}
\toprule
\multicolumn{1}{c}{\textbf{Size}} & 
\multicolumn{1}{c}{\textbf{Ranking}} & 
\multicolumn{1}{c}{\textbf{Context}} &
\multicolumn{3}{c}{\textbf{Code}} & 
\multicolumn{2}{c}{\textbf{Identifier}} &
\multicolumn{1}{c}{} &
\multicolumn{2}{c}{\textbf{No. of CTXs}} \\
\multicolumn{1}{c}{} &
\multicolumn{1}{c}{} &
\multicolumn{1}{c}{} &
\multicolumn{1}{c}{\textbf{EM}} & 
\multicolumn{1}{c}{\textbf{ES}} & 
\multicolumn{1}{c}{\textbf{EM}} & 
\multicolumn{1}{c}{\textbf{P}}& 
\multicolumn{1}{c}{\textbf{R}}& 
\multicolumn{1}{c}{\textbf{F1}} &
\multicolumn{1}{c}{\textbf{AC}}&
\multicolumn{1}{c}{\textbf{RC}}\\
\midrule
\multirow{15}{*}{2048} & \multirow{3}{*}{CodeBERT}   
  & AC & 19.81 & 69.2 & 29.79 & 63.15 & 59.82 & 59.95 & 4.85 & 0 \\
  & & RC & 20.26 & 67.46 & 29.12 & 62.23 & 58.92 & 58.84 & 0 & 3.96  \\
  & & DC & 26.49 & 72.69 & 37.67 & 68.71 & 65.71 & 65.56 & 4.62 & 3.6  \\ \cline{2-11}
  & \multirow{3}{*}{Edit}
    & AC & 19.89 & 69.19 & 29.91 & 63.41 & 60.01 & 60.18 & 4.85 & 0  \\
  & & RC & 20.41 & 67.55 & 29.19 & 62.33 & 58.95 & 58.91 & 0 & 4.07   \\
  & & DC & 27.2 & 72.85 & 38.01 & 68.88 & 65.94 & 65.8 & 4.76 & 3.67  \\ \cline{2-11}
    & \multirow{3}{*}{Jacarrd}
    & AC & 19.89 & 69.23 & 29.91 & 63.24 & 59.89 & 60.04 & 4.85 & 0  \\
  & & RC & 20.71 & 67.61 & 29.46 & 62.5 & 59.19 & 59.1 & 0 & 3.93  \\
  & & DC & 26.87 & 72.9 & 38.2 & 68.89 & 65.94 & 65.78 & 4.72 & 3.44 \\ \cline{2-11}
    & \multirow{3}{*}{Random}
    & AC & 19.96 & 69.28 & 30.09 & 63.29 & 59.9 & 60.06 & 4.85 & 0 \\
  & & RC & 20.26 & 67.46 & 29.19 & 62.22 & 58.91 & 58.85 & 0 & 3.8  \\
  & & DC & 27.13 & 72.45 & 37.67 & 68.36 & 65.37 & 65.24 & 4.41 & 3.43 \\ \cline{2-11}
    & \multirow{3}{*}{UniXcoder}
    & AC & 19.77 & 69.04 & 29.57 & 63.05 & 59.54 & 59.74 & 4.85 & 0  \\
  & & RC& 20.75 & 67.66 & 29.49 & 62.38 & 59.01 & 58.93 & 0 & 3.77  \\
  & & DC & 27.43 & 72.8 & 38.54 & 69.05 & 66.02 & 65.94 & 4.61 & 3.26 \\ \hline
\multirow{15}{*}{4096} & \multirow{3}{*}{CodeBERT}   
  & AC & 19.81 & 69.25 & 29.64 & 63.22 & 59.73 & 59.93 & 4.85 & 0 \\
  & & RC & 21.05 & 68.04 & 30.24 & 62.99 & 59.75 & 59.65 & 0 & 4.31 \\
  & & DC & 27.99 & 73.32 & 38.84 & 69.41 & 66.68 & 66.37 & 4.82 & 4.24 \\ \cline{2-11}
  & \multirow{3}{*}{Edit}
    & AC & 19.81 & 69.12 & 29.79 & 63.4 & 59.99 & 60.14 & 4.85 & 0 \\
  & & RC & 21.09 & 68.11 & 30.32 & 63.16 & 59.87 & 59.79 & 0 & 4.35 \\
  & & DC & 28.03 & 73.49 & 39.06 & 69.62 & 66.82 & 66.59 & 4.84 & 4.29 \\ \cline{2-11}
    & \multirow{3}{*}{Jacarrd}
    & AC & 19.92 & 69.23 & 30.06 & 63.28 & 59.87 & 60.05 & 4.85 & 0 \\
  & & RC & 21.13 & 68.01 & 30.21 & 62.88 & 59.71 & 59.54 & 0 & 4.34 \\
  & & DC & 27.95 & 73.33 & 39.36 & 69.65 & 66.82 & 66.58 & 4.84 & 4.26 \\ \cline{2-11}
    & \multirow{3}{*}{Random}
    & AC & 19.89 & 69.31 & 30.09 & 63.45 & 59.96 & 60.21 & 4.85 & 0 \\
  & & RC & 21.16 & 68.02 & 30.06 & 62.98 & 59.52 & 59.57 & 0 & 4.29  \\
  & & DC & 28.07 & 73.26 & 39.25 & 69.31 & 66.45 & 66.27 & 4.77 & 4.22 \\ \cline{2-11}
    & \multirow{3}{*}{UniXcoder}
    & AC & 19.7 & 69.11 & 29.57 & 62.99 & 59.55 & 59.74 & 4.85 & 0 \\
  & & RC & 21.13 & 68.09 & 30.32 & 63.13 & 59.86 & 59.74 & 0 & 4.3 \\
  & & DC & 27.92 & 73.09 & 39.21 & 69.42 & 66.5 & 66.39 & 4.8 & 4.2 \\ 
\bottomrule
\end{tabular}
\caption{(Continued)Performance with Analogy Context, Rationale Context and Dual Context on Benchmark over DeepSeek-Code-7B.}
\label{tab:allExperiment2}
\end{table}

\clearpage
\section{\GraphName{}}
\label{sec:graph}

The \GraphName{} can construct the dependency relationships between entities in the code and store this information in the form of a multi-directed graph. The \GraphName{} consists of the following main components:

\begin{enumerate}
    \item \textbf{Graph}: This is the core class of the \GraphName{}, which holds the entire graph's data structure of the multi-directed graph and all its nodes and edges.
    
    \item \textbf{Node}: Represents a node in the graph, symbolizing code entities such as modules, classes, functions, or variables.
    
    \item \textbf{Edge}: An object representing a graph edge that defines the relationship between two nodes, characterized by attributes such as an \texttt{EdgeRelation} enumeration value specifying the edge type (e.g., Imports, Calls, etc.), and an optional \texttt{Location} object indicating the edge's position in the source code.
    
    \item \textbf{EdgeRelation}: An enumeration class that lists all possible types of edge relationships.
    
    \item \textbf{NodeType}: An enumeration class that lists all possible types of nodes.
\end{enumerate}

In this design, the \GraphName{} will use the \texttt{graph} attribute to store code entities (\textbf{Node} objects) and their dependency relationships (\textbf{Edge} objects). Each node records its location in the code repository, as well as its type and name. Each edge identifies the type of relationship between two nodes, as well as the location of the relationship in the code.

In the given repository, each \textbf{Node} and \textbf{Edge} is uniquely identified by string representations that are designed to encapsulate their respective attributes. The \textbf{Node} object ID combines the node's name, its type, and its location into a string in the format \texttt{name:type@location}, where the location is represented by the file path and the range of lines and columns it spans. Similarly, the \textbf{Edge} object ID fuses the edge's relation kind with its location, if available, in the format \texttt{relation@location}. Thereby ensuring that each code entity is distinct within the entire repository.

For Python repositories, we employ the tool Jedi \cite{jedi}  to construct the \GraphName, while for other programming languages, alternative tools can be utilized to achieve a similar purpose. The framework is designed with extensibility, allowing for straightforward integration of additional tools to extend support to other programming languages beyond Python.

\begin{figure*}[h]
	\centering
	\includegraphics[width=\linewidth, trim={0 18cm 0 18cm},clip]{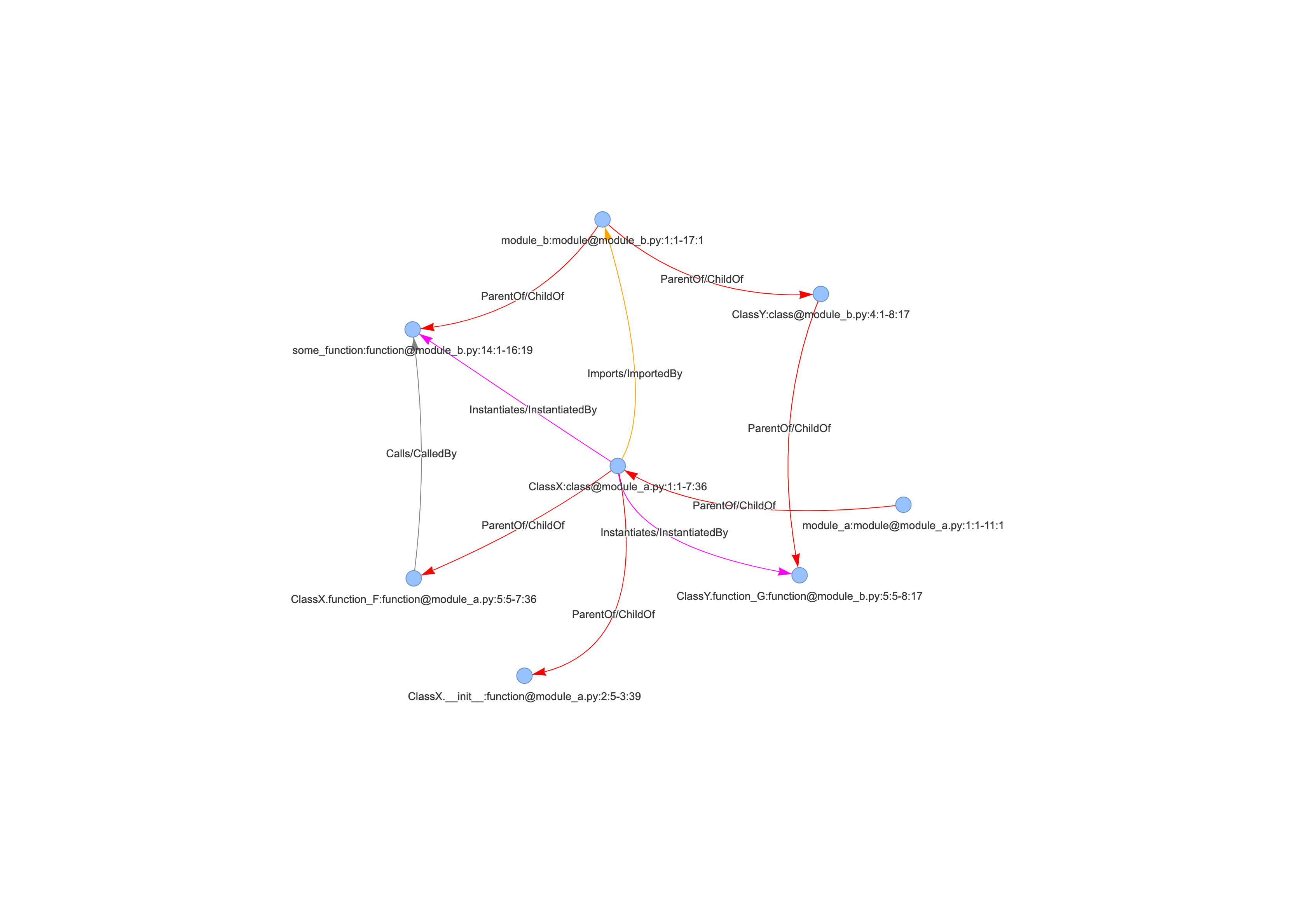}
	\caption{A sample \GraphName{}.}
	\label{fig:sample_graph}
\end{figure*}

To give a simple example, the \cref{fig:sample_graph} represents the knowledge graph depicting relationships such as instantiation of \textit{ClassX} within both \textit{module\_a.py} and \textit{module\_b.py}, the usage of \textit{variable\_V} in \textit{function\_F}, and the import of \textit{ClassX} from \textit{module\_a.py} to \textit{module\_b.py}, all interconnected in a multi-directed graph to capture the complex inter-dependencies in a Python repository which consists of two files: \cref{lst:example1} and \cref{lst:example2}.

\lstinputlisting[language=Python, caption=module\_a.py, label=lst:example1, frame=single]{resources/code_snippets/module\_a.py}
\clearpage
\lstinputlisting[language=Python, caption=module\_b.py, label=lst:example2, frame=single]{resources/code_snippets/module\_b.py}

 \clearpage
\section{Algorithms}

\cref{alg:rationale_context} describes the retrieval of the rationale context for a cursor line within a file by collecting and concatenating the textual content or signatures of relevant nodes that are related by specific edge relations that are located before the cursor line. The nodes and edge relations are constructed by \GraphName as mentioned in \cref{sec:graph}.

\begin{algorithm}[H]
   \caption{Retrieve rationale context by cursor line of file}
   \label{alg:rationale_context}
\begin{algorithmic}[1] 
   \STATE {\bfseries Input:} \GraphName $G$, file path $file\_path$, line number $cursor\_line$
   \STATE {\bfseries Output:} List of strings representing rationale context

   \STATE $node \gets$ Retrieve the innermost node from $G$ at $file\_path$ surrounding $cursor\_line$
   \STATE $related\_edge\_list \gets$ Retrieve related edges from $G$ for $node$, filtering by relations: $Construct$, $BaseClassOf$, $Overrides$, $Calls$, $Instantiates$, $Uses$
   \STATE $module\_node \gets$ Retrieve $MODULE$ type node for $file\_path$ from $G$
   \STATE $importation\_edge\_list \gets$ Retrieve $Imports$ relation edges for $module\_node$ from $G$
   \STATE $all\_edges \gets$ Concatenate $related\_edge\_list$ with $importation\_edge\_list$
   \STATE Initialize empty list $rationale\_context\_list$
   \FORALL{edges $e$ in $all\_edges$}
       \IF{$e$'s out-node is from a different file in the repository \AND $e$'s start line is before $cursor\_line$}
           \STATE $out\_node\_content \gets$ Retrieve content based on $e$'s out-node type:
           \IF{$e$'s out-node is of type $FUNCTION$}
               \STATE $out\_node\_content \gets$ Code text of $e$'s out-node
           \ELSE
               \STATE $out\_node\_content \gets$ Code signature of $e$'s out-node
           \ENDIF
           \STATE Append $out\_node\_content$ to $rationale\_context\_list$
       \ENDIF
   \ENDFOR
   \STATE {\bfseries Return:} $rationale\_context\_list$
\end{algorithmic}
\end{algorithm}
 \clearpage
\section{Throughput-Oriented Efficiency}
\label{sec:effi}

\begin{figure}[!htbp] 
	\centering
	
	\begin{subfigure}[t]{0.48\textwidth} 
		\includegraphics[width=0.9\textwidth]{resources/throughput/1b_a100.pdf}
		\caption{Throughput efficiency of DeepSeek-Coder 1B on NVIDIA A100 GPU with various RAG strategies.}
		\label{fig:throughput_1b_a100}
	\end{subfigure}
	\hfill 
	\begin{subfigure}[t]{0.48\textwidth} 
		\includegraphics[width=0.9\textwidth]{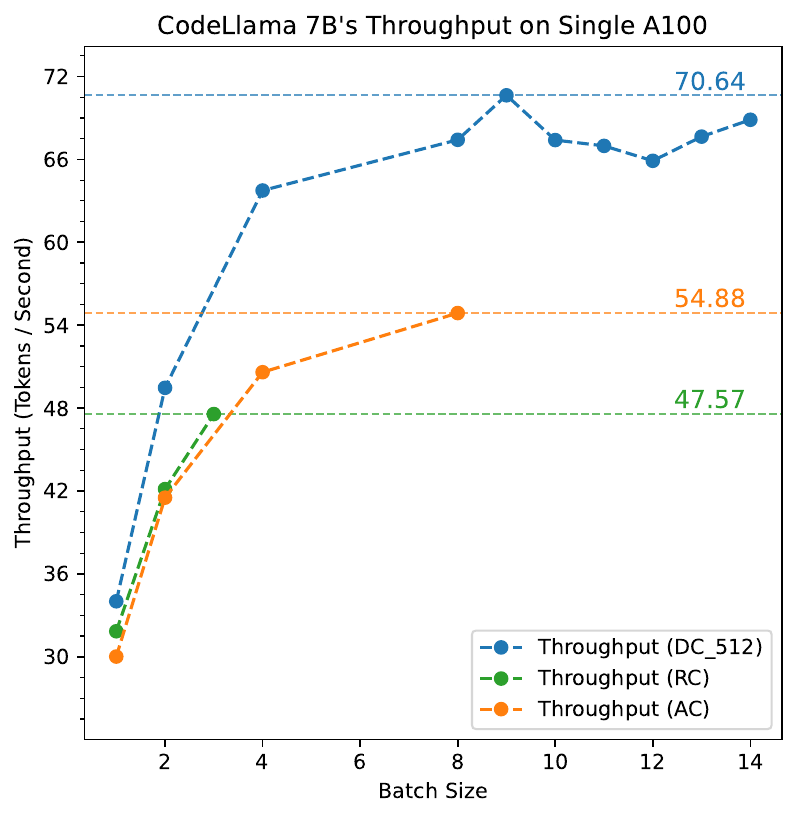}
		\caption{Throughput efficiency of CodeLlama 7B on NVIDIA A100 GPU with various RAG strategies.}
		\label{fig:throughput_7b_a100}
	\end{subfigure}

	\begin{subfigure}[t]{0.48\textwidth}
		\includegraphics[width=0.9\textwidth]{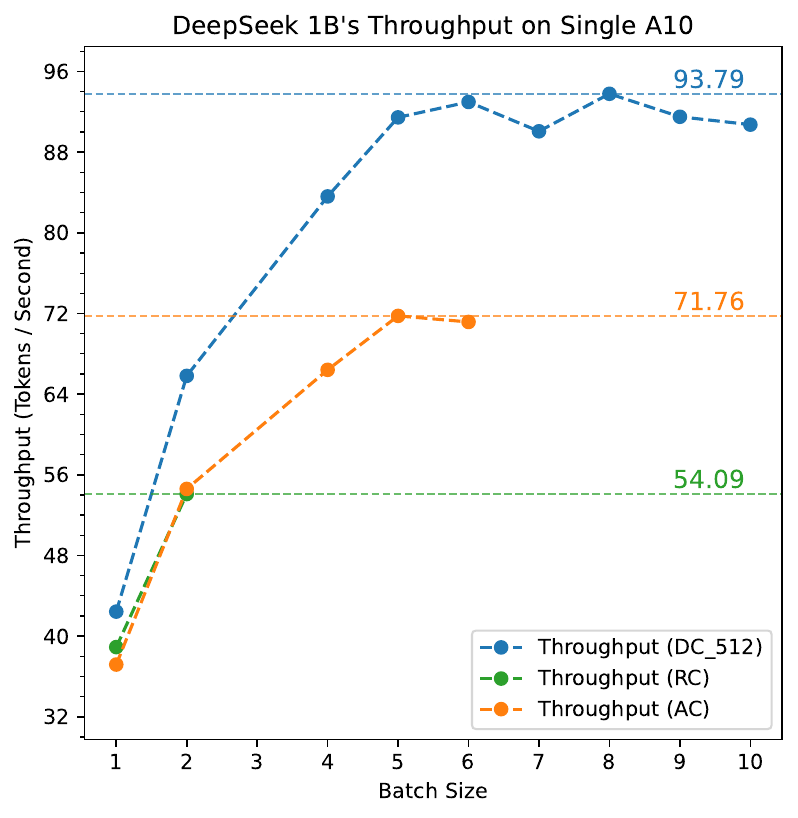}
		\caption{Throughput efficiency of DeepSeek-Coder 1B on NVIDIA A10 GPU with various RAG strategies.}
		\label{fig:throughput_1b_a10}
	\end{subfigure}
	\hfill 
	\begin{subfigure}[t]{0.48\textwidth}
		\includegraphics[width=0.9\textwidth]{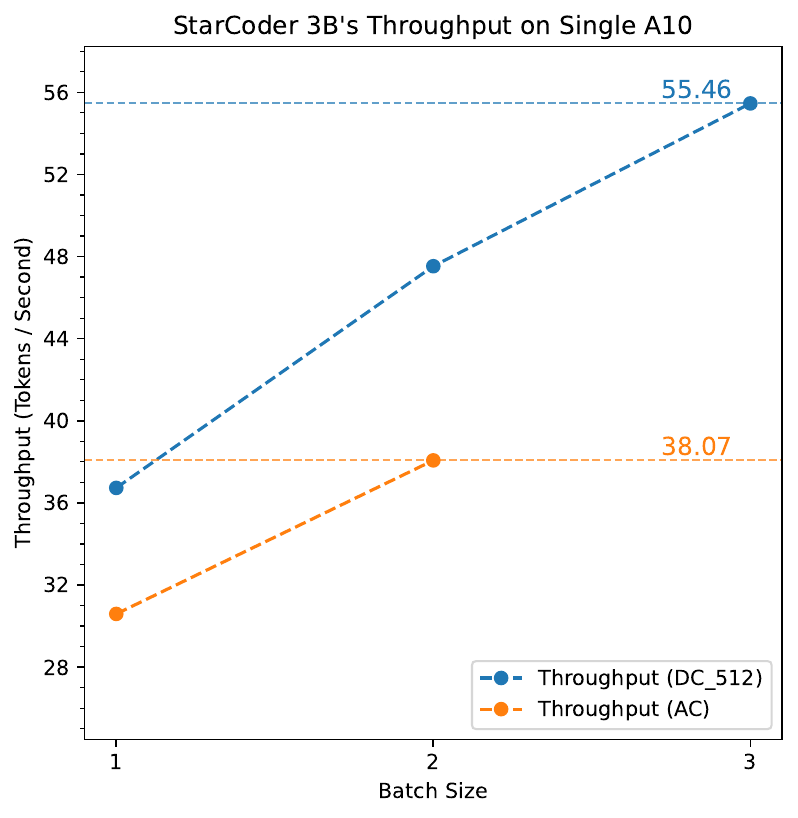}
		\caption{Throughput efficiency of StarCoder 3B on NVIDIA A10 GPU with various RAG strategies.}
		\label{fig:throughput_3b_a10}
	\end{subfigure}

	\caption{Comparative throughput efficiency of different AI model configurations on NVIDIA A100 and A10 GPUs.}
	\label{fig:throughput_all}
\end{figure}

The figures \cref{fig:throughput_1b_a100}, \cref{fig:throughput_7b_a100}, 
\cref{fig:throughput_1b_a10}, and \cref{fig:throughput_3b_a10} demonstrate 
throughput efficiency across various scenarios, highlighting the maximum throughput 
for 1B and 7B models on the A100 GPU, as well as for 1B and 3B models on the A10 GPU.

The DC\_512 configuration achieves 29.2\% to 45.7\% higher peak throughput compared 
to AC, and 56.5\% to 73.4\% more than RC. Notably, the 3b model under the 
RC configuration fails to operate on the NVIDIA A10, underscoring TDC's 
performance advantage.





 \clearpage
\section{Comparison of Different Prompt Construction Strategies}
\label{appendix:ordering}
\begin{figure}[!h]
	\centering
	\includegraphics[width=0.8\linewidth]{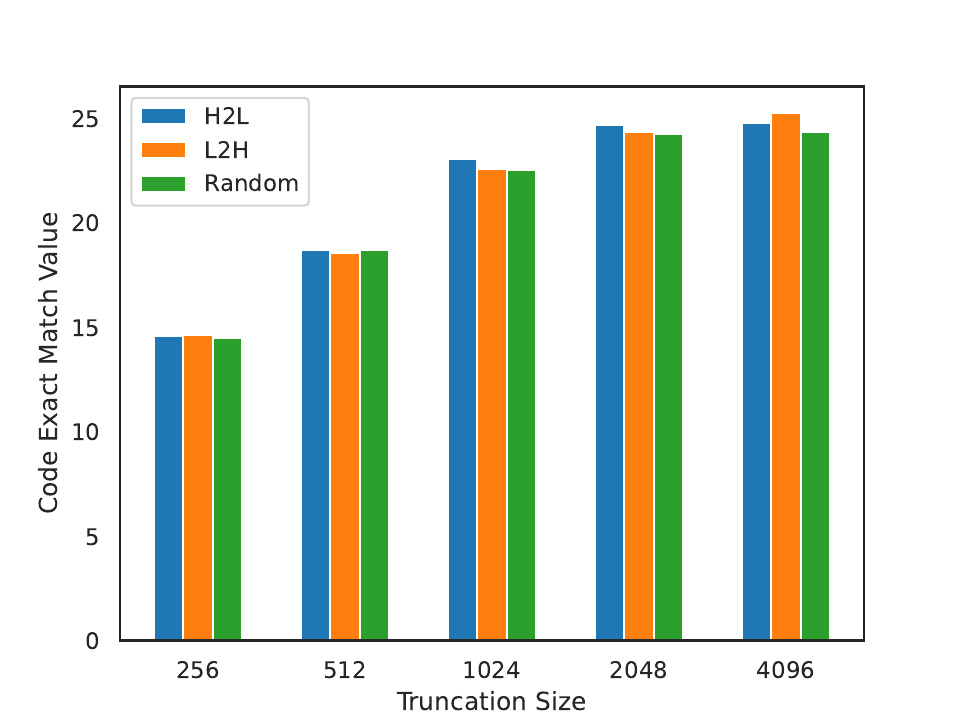}
	\caption{The comparison of different prompt construction strategies on CodeLlama-7B.}
	\label{fig:ordering}
\end{figure}

After selecting DC through a ranking strategy, it becomes essential to understand how to sequence this information for optimal comprehension by code LMs. To this end, we assessed the \textit{Code\_EM} performance of three Prompt Construction strategies: LowToHigh, HighToLow, and Random on the Code Llama-7B model across truncation sizes ranging from 256 to 4096 tokens. \cref{fig:ordering} illustrates that the performances of these strategies are strikingly similar. For instance, at a truncation size of 512, HighToLow and Random marginally outperform LowToHigh, whereas at a truncation size of 4096, LowToHigh shows a slight advantage over HighToLow and Random. At the smallest size of 256, the results of the three strategies are nearly identical. Upon analyzing these outcomes, we deduce that code LMs are relatively indifferent to the order of information presented, suggesting that the strategy of ordering might not be as critical as previously assumed.

 \clearpage
\section{The token length distribution of different context over benchmark.}
\begin{figure}[!h]
	\centering
	\includegraphics[width=0.8\columnwidth]{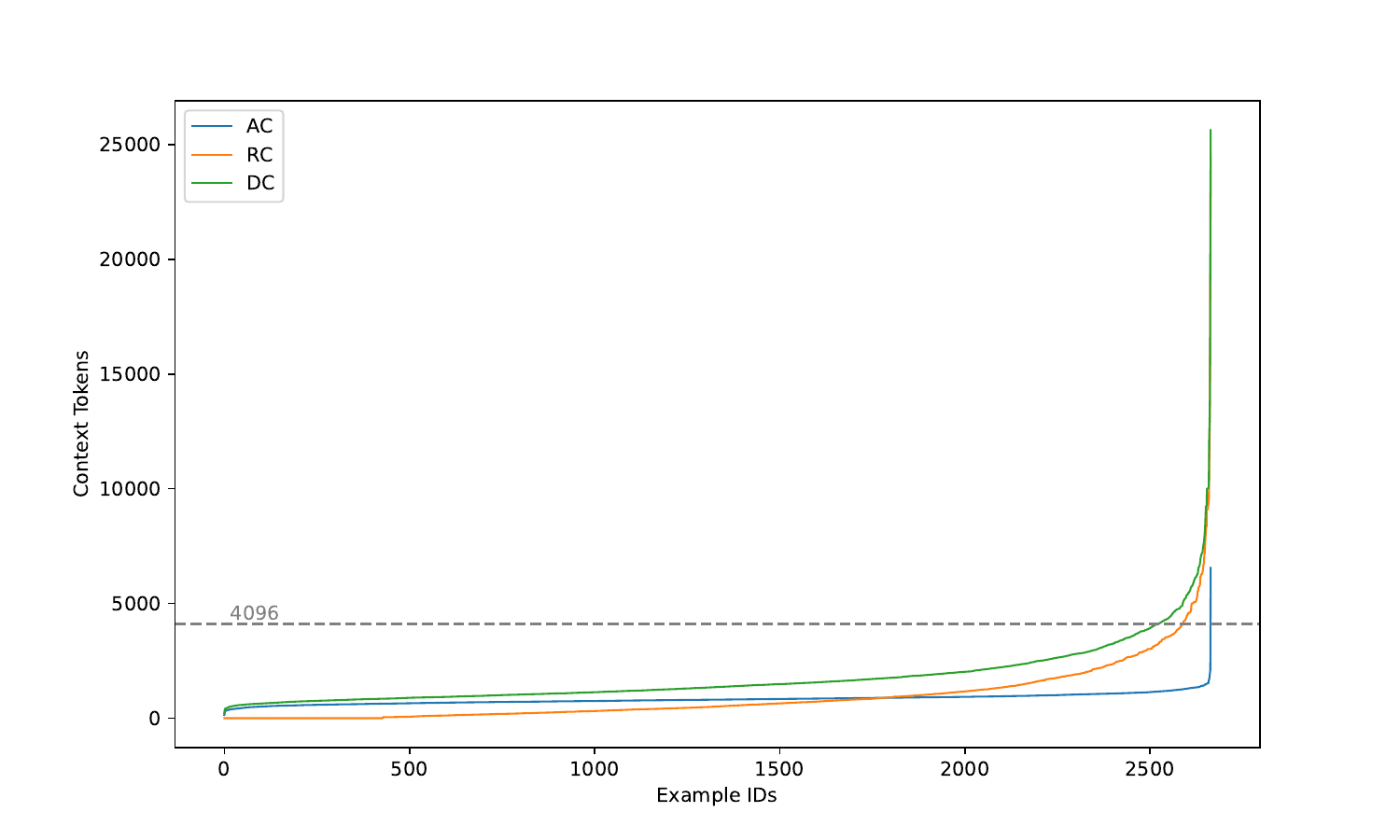}
	\caption{The token length distribution of different context.}
	\label{fig:distribution}
\end{figure}

\section{Convergence of AC, RC and DC over 5 Code LMs}
\label{appendix:convergence}

\begin{figure}[!h]
	\centering
	\includegraphics[width=\columnwidth]{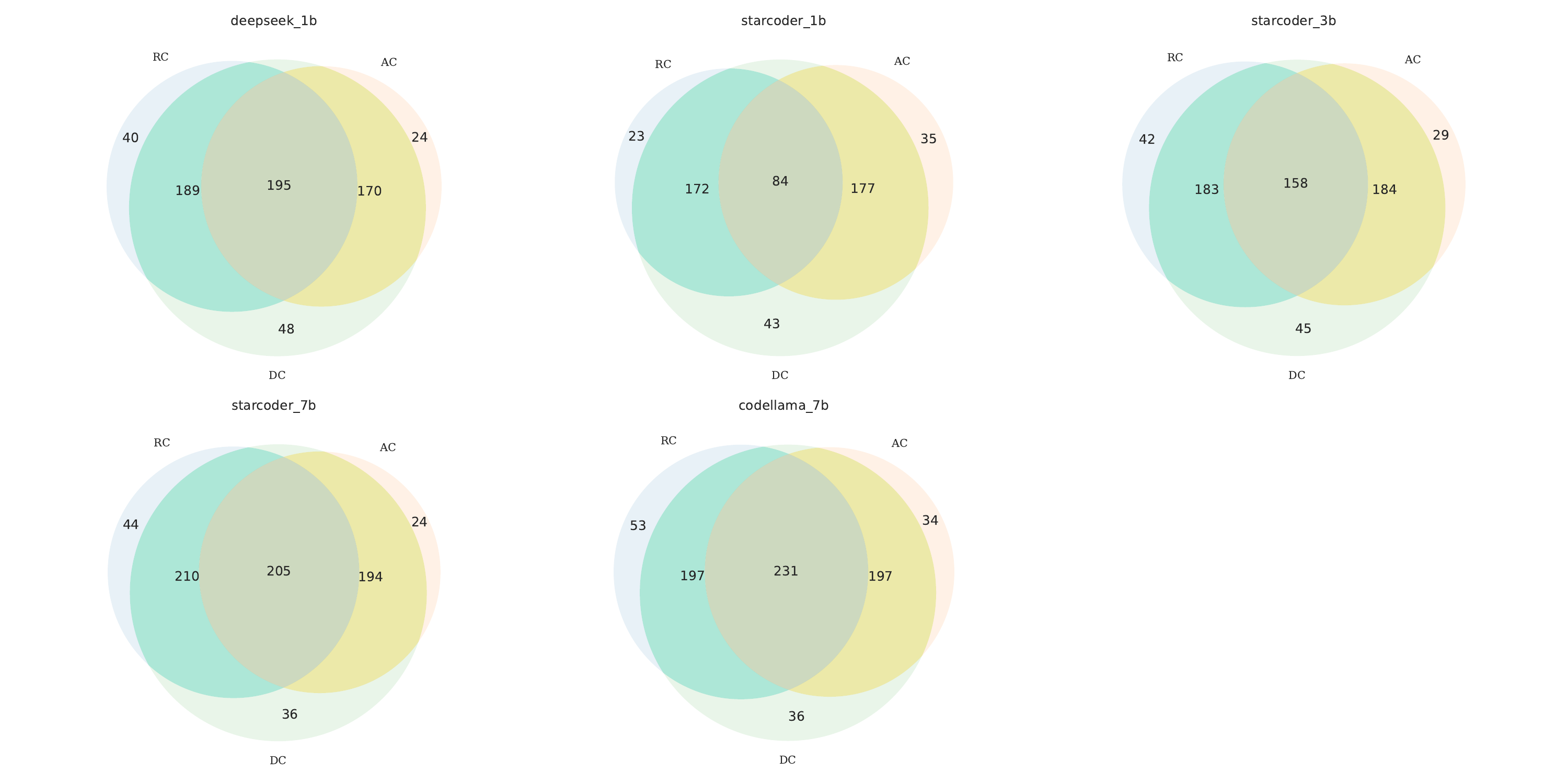}
	\caption{Convergence of AC, RC and DC over other five code LMs. }
	\label{fig:venn_appendix}
\end{figure}

\clearpage
\section{Comparison of Code\_EM values on AC and RC of different Score Functions over Code LMs}
\label{appendix:oracle-function}
\begin{figure}[!h]
	\centering
	\includegraphics[width=0.6\columnwidth]{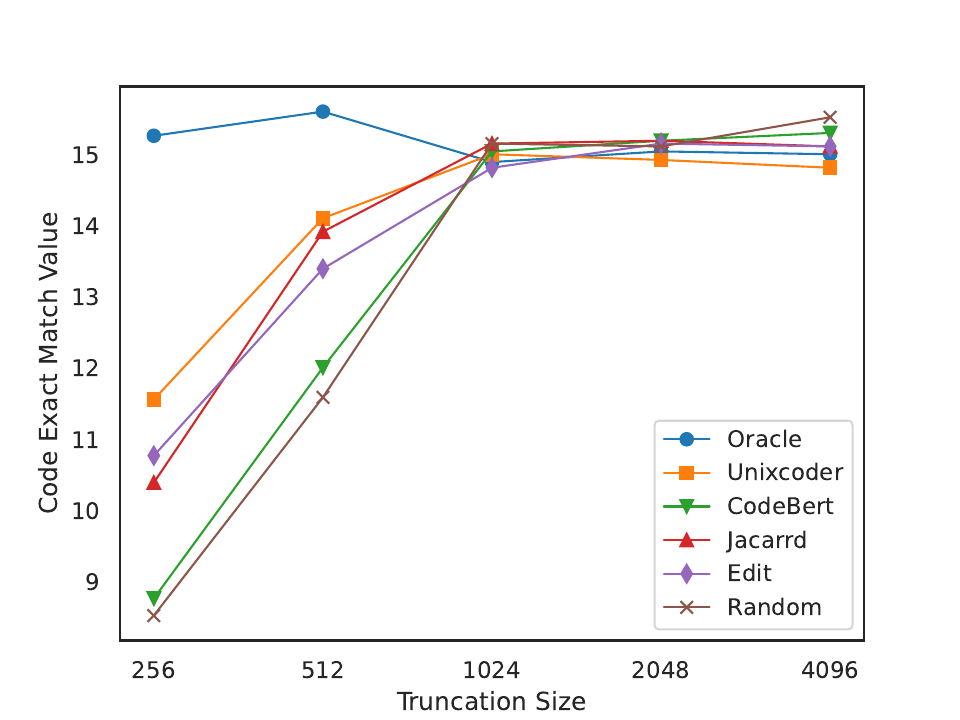}
	\caption{The comparison of Code EM values on all truncation
sizes with AC of different score function.}
	\label{fig:score_function_similar}
\end{figure}

\begin{figure}[!h]
	\centering
	\includegraphics[width=0.6\columnwidth]{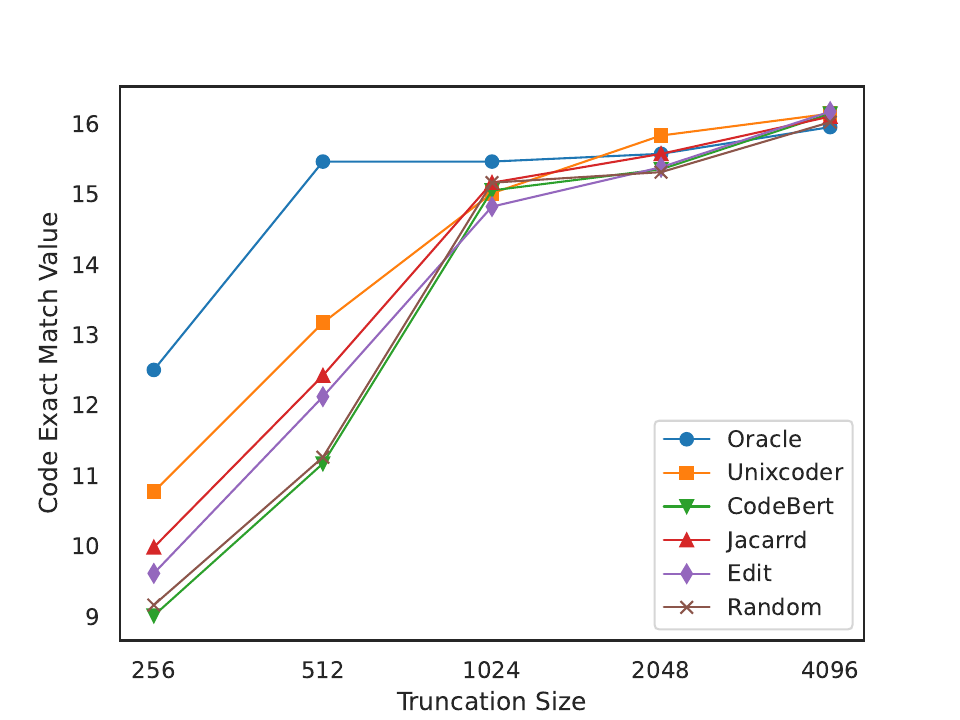}
	\caption{The comparison of Code EM values on all truncation sizes with RC of different score function.}
	\label{fig:score_function_rational}
\end{figure}

\clearpage
\section{A Practical Prompt Example}
\label{appendix:prompt-example}
\begin{figure}[!h]
	\centering
	\includegraphics[width=\columnwidth]{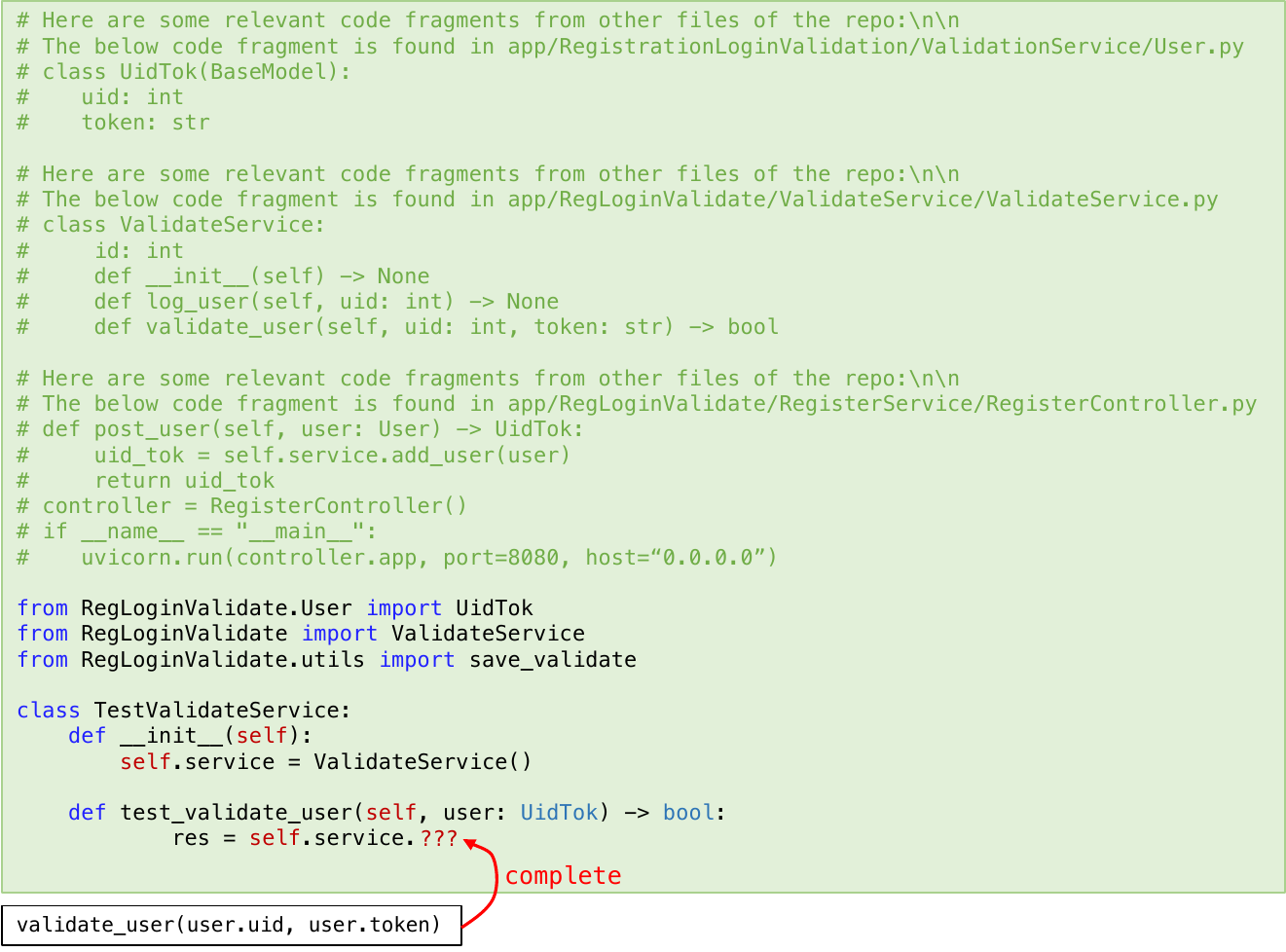}
	\caption{A prompt example.}
	\label{fig:prompt_example}
\end{figure}

\end{document}